\documentclass[10pt]{article}

\usepackage{amssymb,amsmath,amsthm,mathrsfs} 

\usepackage[authoryear,sort&compress]{natbib}
\usepackage{colordvi,color,pspicture}
\usepackage{rotating}

\usepackage{setspace}

\newcommand{\X}{\mathcal{X}}
\newcommand{\U}{\mathcal{U}}

\newcommand{\E}{\mathbf{E}}
\newcommand{\Var}{\mathbf{Var}}
\newcommand{\Cov}{\mathbf{Cov}}

\newcommand{\al}{\alpha}

\newcommand{\bh}{\widehat{\beta}}

\theoremstyle{plain}
\newtheorem{theorem}{Theorem}[section]

\theoremstyle{definition}

\theoremstyle{remark}

\newtheorem{remark}[theorem]{Remark}

\begin{document}

\title{Technical Report \# KU-EC-08-5:\\
QR-Adjustment for Clustering Tests Based on Nearest Neighbor Contingency Tables}
\author{
Elvan Ceyhan
\thanks{Department of Mathematics, Ko\c{c} University, 34450 Sar{\i}yer, Istanbul, Turkey}
}
\date{\today}
\maketitle

\begin{singlespace}
\begin{abstract}
The spatial interaction between two or more classes of points may
cause spatial clustering patterns such as segregation or
association, which can be tested using a nearest neighbor
contingency table (NNCT).
A NNCT is constructed using the
frequencies of class types of points in nearest neighbor (NN) pairs.
For tests based on NNCTs (i.e., NNCT-tests), the null pattern
is either complete spatial randomness (CSR) of the points from two
or more classes (called CSR independence) or
random labeling (RL).
The RL pattern implies that the locations of
the points in the study region are fixed,
while the CSR independence pattern implies that they are random.
The distributions of the
NNCT-test statistics depend on the number of reflexive NNs (denoted
by $R$) and the number of shared NNs (denoted by $Q$), both of which
depend on the allocation of the points.
Hence $Q$ and $R$ are fixed
quantities under RL, but random variables under CSR independence.
However given the difficulty in calculating the expected values of
$Q$ and $R$ under CSR independence,
one can use their observed values in NNCT analysis,
which makes the distributions of the
NNCT-test statistics conditional on $Q$ and $R$ under CSR independence.
In this article, I use the empirically estimated expected
values of $Q$ and $R$ under CSR independence pattern to remove
the conditioning of NNCT-tests (such a correction is called the
\emph{QR-adjustment}, henceforth).
I present a Monte Carlo simulation
study to compare the conditional NNCT-tests (i.e., tests with the observed values of $Q$
and $R$ are used) and unconditional NNCT-tests (i.e., empirically QR-adjusted tests)
under CSR independence and segregation and association alternatives.
I demonstrate that QR-adjustment does not significantly improve
the empirical size estimates under CSR independence
and power estimates under segregation or association alternatives.
For illustrative purposes, I apply the
conditional and empirically corrected tests on two example data sets.
\end{abstract}

\noindent
{\small {\it Keywords:}
Association; complete spatial randomness;
conditional test; nearest neighbor contingency table;
random labeling; spatial clustering; spatial pattern; segregation

\vspace{.25 in}

$^*$corresponding author.\\
\indent {\it e-mail:} elceyhan@ku.edu.tr (E.~Ceyhan) }

\end{singlespace}

\newpage

\section{Introduction}
\label{sec:intro}
Spatial patterns have been studied extensively
and have important implications in many fields
such as epidemiology, population biology, and ecology.
It is of practical interest to study the univariate spatial
patterns (i.e., patterns of only one class)
as well as multivariate patterns (i.e., patterns of multiple classes)
(\cite{pielou:1961}, \cite{whipple:1980}, and \cite{dixon:1994,dixon:NNCTEco2002}).
For convenience and generality,
I call the different types of points as ``classes",
but a class label can stand for any characteristic of a measurement
at a particular location.
For example, the spatial segregation pattern
has been investigated for \emph{species} (\cite{diggle:2003}),
\emph{age classes} of plants (\cite{hamill:1986}),
\emph{fish species} (\cite{herler:2005}),
and \emph{sexes} of dioecious plants (\cite{nanami:1999}).
Many of the epidemiological applications are for a two-class system of
case and control labels (\cite{waller:2004}).

In this article, for simplicity, I describe the spatial point patterns for two
classes only; the extension to multi-class case is straightforward.
The null pattern is usually one of the two (random) pattern
types: \emph{complete spatial randomness} (CSR)
of two or more classes or \emph{random labeling} (RL)
of a set of fixed points with two classes.
That is, when the points from each
class are assumed to be uniformly distributed over the region of interest,
the the null hypothesis is the CSR of points from two classes.
This type of CSR pattern is also referred to as ``population independence"
in literature (\cite{goreaud:2003}).
In the univariate spatial analysis,
CSR refers to the pattern in which locations of
points from a \emph{single} class are random over the study area.
To distinguish the CSR of points from two-classes and CSR of points from one class,
I call the former as ``CSR independence" and the latter as ``CSR", henceforth.
Note that CSR independence is equivalent
to the case that RL procedure is applied to a
given set of points from a CSR pattern
in the sense that after points are generated uniformly
in the region, the class labels are assigned randomly.
When only the labeling of a set of fixed points
(the allocation of the points could be regular, aggregated,
or clustered, or of lattice type)
is random, the null hypothesis is the RL pattern.

Many tests of spatial segregation have been developed in literature (\cite{orton:1982}).
These include comparison of Ripley's $K(t)$ or $L(t)$ functions (\cite{ripley:2004}),
comparison of nearest neighbor (NN) distances (\cite{diggle:2003}, \cite{cuzick:1990}),
and analysis of nearest neighbor contingency tables (\cite{pielou:1961}, \cite{meagher:1980}).
Nearest neighbor contingency tables (NNCTs) are constructed using the frequencies of
classes of points in NN pairs.
\cite{kulldorff:2006} provides an extensive review of tests of spatial randomness that
adjust for an inhomogeneity of the densities of the underlying populations.
\cite{pielou:1961} proposed various tests and (\cite{dixon:1994}) introduced an
overall test of segregation, cell-, and class-specific tests based on NNCTs
for the two-class case  and extended his tests to
multi-class case (\cite{dixon:NNCTEco2002}).
These tests based on NNCTs (i.e., NNCT-tests)
were designed for testing the RL of points.
\cite{pielou:1961} used the usual Pearson's $\chi^2$-test of
independence for detecting the segregation of the two classes.
Due to the ease in computation and interpretation,
Pielou's test of segregation is frequently used for both CSR independence and RL patterns.
However it has been shown that Pielou's test is not appropriate for testing RL
(\cite{meagher:1980}, \cite{dixon:1994}).
\cite{dixon:1994} derived the appropriate (asymptotic) sampling
distribution of cell counts using Moran join count statistics (\cite{moran:1948}) and
hence the appropriate test which also has a $\chi^2$-distribution (\cite{dixon:1994}).
For the two-class case, \cite{ceyhan:overall} compared these tests,
extended the tests for testing CSR independence,
and demonstrated that Pielou's tests are only appropriate for
a random sample of (base, NN) pairs.
Furthermore, \cite{ceyhan:corrected} proposed three new overall segregation tests.
Since Pielou's test is not appropriate,
NNCT-tests only refer to Dixon's overall test and the three new segregation tests
proposed by \cite{ceyhan:corrected}.
However the distributions of the NNCT-test statistics depend on the
number of reflexive NNs (denoted by $R$) and the number of shared NNs (denoted by $Q$),
both of which depend on the allocation of the points.
Hence $Q$ and $R$ are fixed under RL,
but random under CSR independence.
But expectations of $Q$ and $R$ seem to be not available analytically under the
CSR independence, so their observed values
were used by \cite{ceyhan:corrected}.
In this article, I replace the expectations of $Q$ and $R$
by their empirical estimates under CSR independence.
Such a correction for removing the conditional nature of NNCT-tests
is called ``QR-adjustment", henceforth.

The NNCT-tests are designed for testing a more general null hypothesis, namely,
$H_o:$ \emph{randomness in the NN structure},
which usually results from CSR independence or RL.
The distinction between CSR independence and RL is very important
when defining the appropriate null model for each empirical case,
i.e., the null model depends on the particular context.
\cite{goreaud:2003} discuss the differences between these two null hypotheses
and demonstrate that the misinterpretation is very common.
They assert that under CSR independence the (locations of the points from)
two classes are \emph{a priori} the result of different processes
(e.g., individuals of different species or age cohorts),
whereas under RL some processes affect \emph{a posteriori}
the individuals of a single population
(e.g., diseased versus non-diseased individuals of a single species).
Notice that although CSR independence and RL are not same,
they lead to the same null model (i.e., randomness in NN structure) for NNCT-tests,
since a NNCT does not require spatially-explicit information.

I consider two major types of (bivariate) spatial clustering patterns, namely,
\emph{association} and \emph{segregation} as alternative patterns.
{\em Association} occurs if the NN of an individual is more
likely to be from another class.
{\em Segregation} occurs if the
NN of an individual is more likely to be of the same class as the individual;
i.e., the members of the same class tend to be clumped or clustered
(see, e.g., \cite{pielou:1961}).
For more detail on these alternative patterns, see (\cite{ceyhan:corrected}).
I assess the effects of QR-adjustment on the size of the NNCT-tests under CSR independence
and on the power of the tests under the segregation or association alternatives
by an extensive Monte Carlo study.

Throughout the article I adopt the convention that
random quantities are denoted by capital letters,
while fixed quantities are denoted by lower case letters.
I describe the construction of NNCTs in Section \ref{sec:NNCT},
provide Dixon's tests in Sections
\ref{sec:Dixon-cell-spec} and \ref{sec:dixon-overall},
empirical significance levels of the tests in Section \ref{sec:emp-sign-level},
two illustrative examples in Section \ref{sec:examples},
and discussion and conclusions in Section \ref{sec:disc-conc}.

\section{Nearest Neighbor Contingency Tables and Related Tests}

\subsection{Construction of the Nearest Neighbor Contingency Tables}
\label{sec:NNCT}
NNCTs are constructed using the NN frequencies of classes.
I describe the construction of NNCTs for two classes; extension to multi-class
case is straightforward.
Consider two classes with labels $\{1,2\}$.
Let $N_i$ be the number of points from class $i$ for $i \in \{1,2\}$ and
$n$ be the total sample size, so $n=N_1+N_2$.
If I record the class
of each point and the class of its NN, the NN
relationships fall into four distinct categories:
$(1,1),\,(1,2);\,(2,1),\,(2,2)$ where in cell $(i,j)$, class $i$ is
the \emph{base class}, while class $j$ is the class of its
\emph{NN}.
That is, the $n$ points constitute $n$
(base, NN) pairs.
Then each pair can be categorized with respect to
the base label (row categories) and NN label (column categories).
Denoting $N_{ij}$ as the frequency of cell $(i,j)$ for $i,j \in
\{1,2\}$, I obtain the NNCT in Table \ref{tab:NNCT-2x2} where $C_j$
is the sum of column $j$; i.e., number of times class $j$ points
serve as NNs for $j \in \{1,2\}$.
Furthermore, $N_{ij}$ is the cell count for
cell $(i,j)$ that is the sum of all (base, NN) pairs each of which
has label $(i,j)$.
Note also that
$n=\sum_{i,j}N_{ij}$; $n_i=\sum_{j=1}^2\, N_{ij}$; and
$C_j=\sum_{i=1}^2\, N_{ij}$.
By construction, if $N_{ij}$ is larger (smaller) than expected,
then class $j$ serves as NN more (less) to class $i$ than expected,
which implies (lack of) segregation if $i=j$ and (lack of) association of class $j$
with class $i$ if $i\not=j$.
Hence, column sums, cell counts are random, while row sums and the overall sum are fixed
quantities in a NNCT.

\begin{table}[ht]
\centering
\begin{tabular}{cc|cc|c}
\multicolumn{2}{c}{}& \multicolumn{2}{c}{NN class}& \\
\multicolumn{2}{c|}{}& class 1 &  class 2 & sum  \\
\hline
&class 1 &    $N_{11}$  &   $N_{12}$  &   $n_1$  \\
\raisebox{1.5ex}[0pt]{base class}
&class 2 &    $N_{21}$ &  $N_{22}$    &   $n_2$  \\
\hline
& sum    &    $C_1$   & $C_2$         &   $n$  \\
\end{tabular}
\caption{
\label{tab:NNCT-2x2}
The NNCT for two classes.}
\end{table}

Observe that, under segregation, the diagonal entries,
$N_{ii}$ for $i=1,2$, tend to be larger than expected; under
association, the off-diagonals tend to be larger than expected.
The general alternative is that some cell counts are different
than expected under CSR independence or RL.

In the two-class case, \cite{pielou:1961} used Pearson's $\chi^2$-test
of independence to detect any deviation from CSR independence or RL.
But, under CSR independence or RL, this test is liberal, i.e.,
has larger size than the nominal level (\cite{ceyhan:overall}),
hence not considered in this article.
\cite{dixon:1994} proposed a series of tests for segregation based on NNCTs.
He first devised four cell-specific tests in the two-class case,
and then combined them to form an overall test.
For his tests, the probability of an individual from
class $j$ serving as a NN of an individual from class $i$
depends only on the class sizes (i.e., row sums),
but not the total number of times class $j$ serves as NNs (i.e., column sums).

\subsection{Dixon's Cell-Specific Tests}
\label{sec:Dixon-cell-spec}
The level of segregation is estimated by
comparing the observed cell counts to the expected cell counts
under RL of points that are fixed.
Dixon demonstrates that under RL,
one can write down the cell frequencies as Moran join count
statistics (\cite{moran:1948}).
He then derives the means, variances, and covariances of
the cell counts (frequencies) in a NNCT (\cite{dixon:1994,dixon:NNCTEco2002}).

The null hypothesis under RL is given by
\begin{equation}
\label{eqn:Exp[Nij]}
H_o:\,\mathbf {E}[N_{ij}]=
\begin{cases}
\frac{n_i(n_i-1)}{(n-1)} & \text{if $i=j$,}\\
\frac{n_i\,n_j}{(n-1)}     & \text{if $i \not= j$.}
\end{cases}
\end{equation}
Observe that the expected cell counts depend only on the size of
each class (i.e., row sums), but not on column sums.

The cell-specific test statistics suggested by Dixon are given by
\begin{equation}
\label{eqn:dixon-Zij}
Z^D_{ij}=\frac{N_{ij}-\mathbf {E}[N_{ij}]}{\sqrt{\Var[N_{ij}]}},
\end{equation}
where
{\small
\begin{multline}
\label{eqn:VarNij}
\Var[N_{ij}]=
\begin{cases}
(n+R)\,p_{ii}+(2\,n-2\,R+Q)\,p_{iii}+(n^2-3\,n-Q+R)\,p_{iiii}-(n\,p_{ii})^2 & \text{if $i=j$,}\\
n\,p_{ij}+Q\,p_{iij}+(n^2-3\,n-Q+R)\,p_{iijj} -(n\,p_{ij})^2         & \text{if $i \not= j$,}
\end{cases}
\end{multline}
}
with $p_{xx}$, $p_{xxx}$, and $p_{xxxx}$
are the probabilities that a randomly picked pair,
triplet, or quartet of points, respectively, are the indicated classes and
are given by
\begin{align}
\label{eqn:probs}
p_{ii}&=\frac{n_i\,(n_i-1)}{n\,(n-1)},  & p_{ij}&=\frac{n_i\,n_j}{n\,(n-1)},\nonumber\\
p_{iii}&=\frac{n_i\,(n_i-1)\,(n_i-2)}{n\,(n-1)\,(n-2)}, &
p_{iij}&=\frac{n_i\,(n_i-1)\,n_j}{n\,(n-1)\,(n-2)},\\
 p_{iijj}&=\frac{n_i\,(n_i-1)\,n_j\,(n_j-1)}{n\,(n-1)\,(n-2)\,(n-3)},&
p_{iiii}&=\frac{n_i\,(n_i-1)\,(n_i-2)\,(n_i-3)}{n\,(n-1)\,(n-2)\,(n-3)}.\nonumber
\end{align}
Furthermore, $Q$ is the number of points with shared  NNs,
which occur when two or more
points share a NN and
$R$ is twice the number of reflexive pairs.
Then $Q=2\,(Q_2+3\,Q_3+6\,Q_4+10\,Q_5+15\,Q_6)$
where $Q_k$ is the number of points that serve
as a NN to other points $k$ times.
One-sided and two-sided tests are possible for each cell $(i,j)$
using the asymptotic normal approximation of $Z^D_{ij}$ given in
Equation (\ref{eqn:dixon-Zij}) (\cite{dixon:1994}).
The test in Equation \eqref{eqn:dixon-Zij} is the same as
Dixon's $Z_{AA}$ when $i=j=1$;
same as $Z_{BB}$ when $i=j=2$ (\cite{dixon:1994}).
Note also that in Equation \eqref{eqn:dixon-Zij} four different tests are defined
as there are four cells and each is testing the deviation from the null case
in the respective cell.
These four tests are combined and
used in defining an overall test of segregation in Section \ref{sec:dixon-overall}.

Under CSR independence, the null hypothesis, the test statistics, and
the variances are as in the RL case for the cell-specific tests,
except for the fact that the variances are conditional on $Q$ and $R$.

\subsection{The Status of $Q$ and $R$ under CSR Independence and RL}
\label{sec:QandR}
Note the difference  in status of the variables $Q$ and $R$ under CSR independence and RL models.
Under RL, $Q$ and $R$ are fixed quantities;
while under CSR independence, they are random.
The quantities given in Equations \eqref{eqn:Exp[Nij]}, \eqref{eqn:VarNij},
and all the quantities depending on these expectations also depend on $Q$ and $R$.
Hence these expressions are appropriate under the RL pattern.
Under CSR independence pattern they are conditional variances and
covariances obtained by using the observed values of $Q$ and $R$.
The unconditional variances and covariances can be obtained
by replacing $Q$ and $R$ with their expectations.

\begin{figure}[ht]
\centering
\rotatebox{-90}{ \resizebox{2.3 in}{!}{\includegraphics{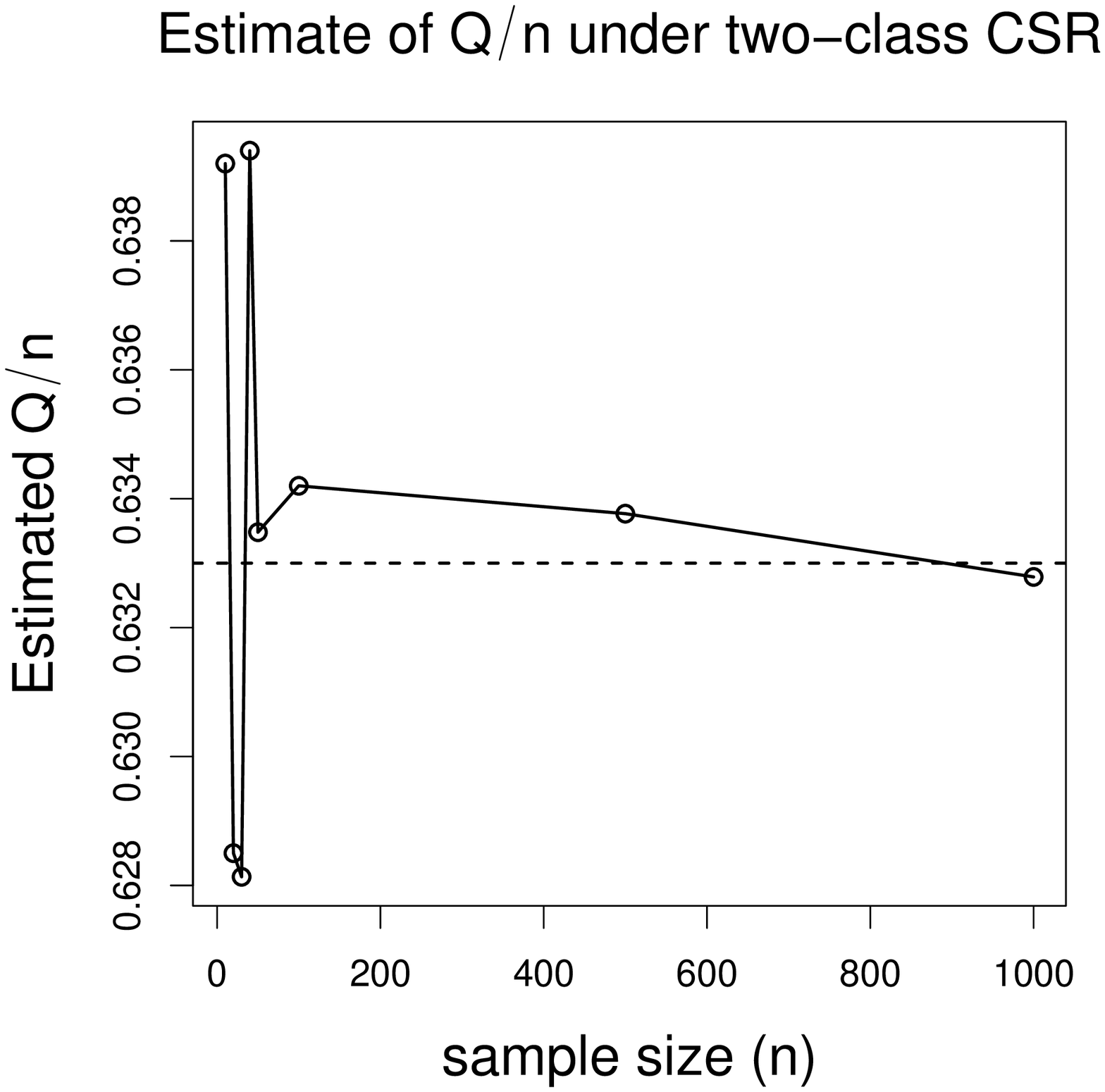} }}
\rotatebox{-90}{ \resizebox{2.3 in}{!}{\includegraphics{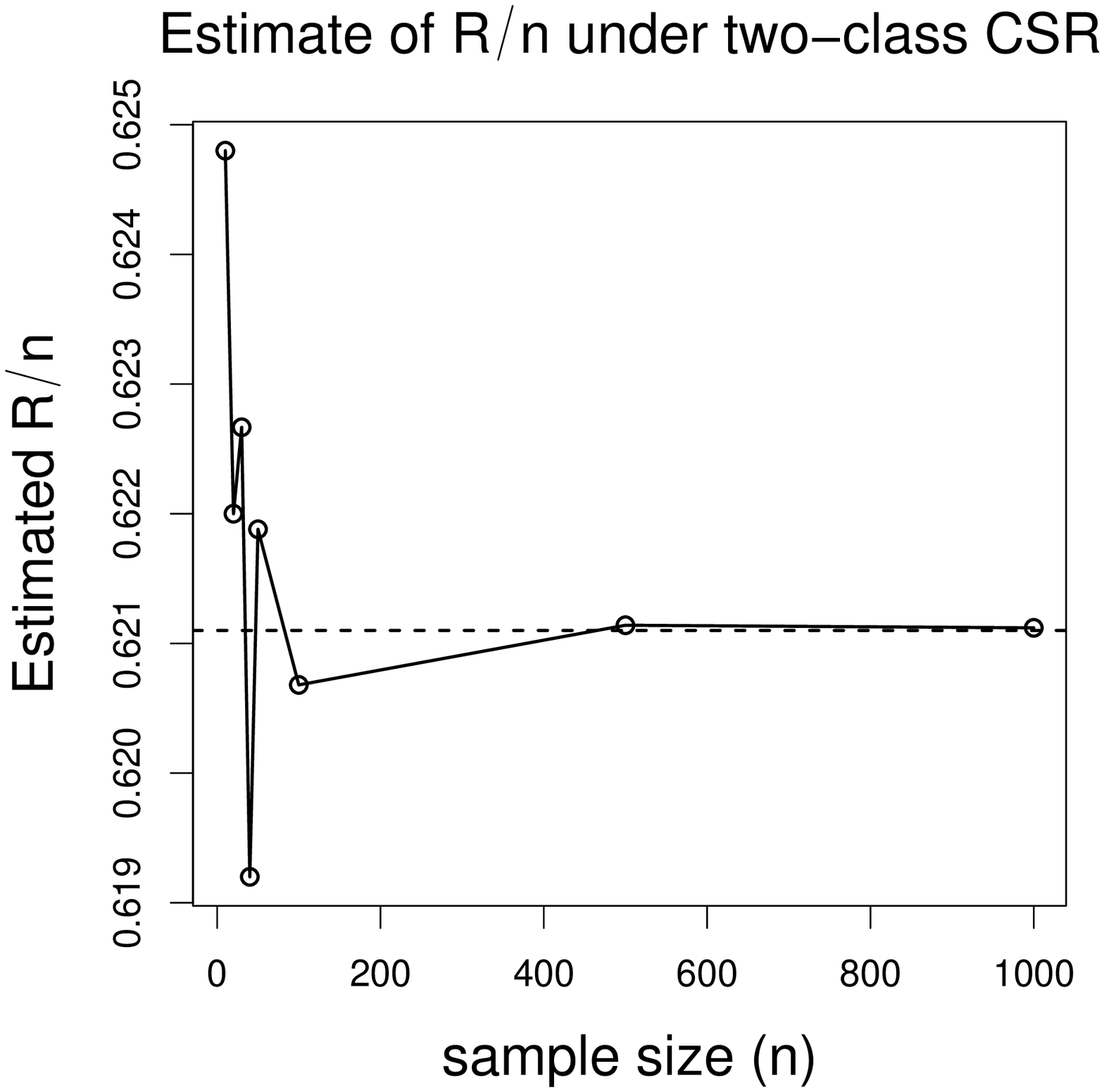} }}
\caption{
\label{fig:Q-R-estimate}
Plotted are the empirically estimated expectations $\E[Q/n]$ (left)
and $\E[R/n]$ (right) as a function of total sample size $n$.
}
\end{figure}

Unfortunately, given the difficulty of calculating the
expectations of $Q$ and $R$ under CSR independence,
\cite{ceyhan:corrected} employed the conditional variances and covariances
(i.e., the variances and covariances for which observed $Q$ and $R$ values
are used) even when assessing their behavior under CSR independence pattern.
Alternatively, I can estimate the values of $Q$ and $R$ empirically as follows.
I generate $n \in\{10,20,30,40,50,100,500,1000\}$
points that are iid (independently and identically distributed)
from $\U((0,1)\times (0,1))$, the uniform distribution on the unit square.
I repeat this procedure $N_{mc}=1000000$ times.
At each Monte Carlo replication, I calculate $Q$ and $R$ values,
and record the ratios $Q/n$ and $R/n$.
I plot these ratios in Figure \ref{fig:Q-R-estimate} as a function of sample size $n$.
Observe that the ratios seem to converge as $n$ increases.
For homogeneous planar Poisson pattern,
I have $\E[Q/n] \approx .6327860$ and $\E[R/n] \approx 0.6211200$.
Hence, I replace $Q$ and $R$ by $0.63\,n$ and $0.62\,n$, respectively,
to obtain the QR-adjusted variances and covariances.

\subsection{Dixon's Overall Segregation Test}
\label{sec:dixon-overall}
Dixon's overall test of segregation tests the hypothesis that expected
cell counts in the NNCT are as in Equation \eqref{eqn:Exp[Nij]}.
In the two-class case, he calculates
$Z_{ii}=(N_{ii}-\E[N_{ii}])\big/\sqrt{\Var[N_{ii}]}$ for both $i \in
\{1,2\}$ and combines these test statistics into a statistic
that is asymptotically distributed as $\chi^2_2$ under RL (\cite{dixon:1994}).
The suggested test statistic is given by
{\small
\begin{multline}
\label{eqn:dix-chisq-2x2}
C=\mathbf{Y}'\Sigma^{-1}\mathbf{Y}=
\left[
\begin{array}{c}
N_{11}-\E[N_{11}] \\
N_{22}-\E[N_{22}]
\end{array}
\right]'
\left[
\begin{array}{cc}
\Var[N_{11}] & \Cov[N_{11},N_{22}] \\
\Cov[N_{11},N_{22}] & \Var[N_{22}] \\
\end{array}
\right]^{-1}
\left[
\begin{array}{c}
N_{11}-\E[N_{11}] \\
N_{22}-\E[N_{22}]
\end{array}
\right],
\end{multline}
}
where $\E[N_{ii}]$ are as in Equation (\ref{eqn:Exp[Nij]}),
$\Var[N_{ii}]$ are as in Equation (\ref{eqn:VarNij}),
and
\begin{equation}
\label{eqn:cov-N11-N22}
\Cov[N_{11},\,N_{22}]=(n^2-3\,n-Q+R)\,p_{1122}-n^2\,p_{11}\,p_{22}.
\end{equation}
Dixon's $C$ statistic given in Equation \eqref{eqn:dix-chisq-2x2} can also be written as
$$C=\frac{Z_{AA}^2+Z_{BB}^2-2rZ_{AA}Z_{BB}}{1-r^2},$$
where $r=\Cov[N_{11},N_{22}] \Big /\sqrt{\Var[N_{11}]\Var[N_{22}]}$ (\cite{dixon:1994}).

Under CSR independence, the expected values, variances and covariances are as in the RL case.
However, the variance and covariance terms include $Q$ and $R$ which are random
under CSR independence and fixed under RL.
Hence Dixon's test statistic $C$ asymptotically has a $\chi^2_1$-distribution
under CSR independence conditional on $Q$ and $R$.
Replacing $Q$ and $R$ by their empirical estimates given in Section \ref{sec:QandR},
I obtain the QR-adjusted version of Dixon's test which is denoted by $C_{qr}$.

\subsection{Version I of the New Segregation Tests}
\label{sec:seg-test-I}
\cite{ceyhan:corrected} proposed tests based on the correct sampling distribution of
the cell counts in a NNCT under CSR independence or RL.
In defining the new segregation or clustering tests, I follow
a track similar to that of Dixon's (\cite{dixon:1994})
where he defines a cell-specific test statistic for each cell and then
combines these four tests into an overall test.

For cell $(i,j)$, let
\begin{equation}
\label{eqn:seg-test-I}
T^I_{ij}=N_{ij}-\frac{n_i\,C_j}{n}~~~ \text{  and then let  }~~~ N^I_{ij}=\frac{T^I_{ij}}{\sqrt{n_i\,c_j/n}}=
\frac{\left( N_{ij}-n_i\,c_j/n \right)}{ \sqrt{n_i\,c_j/n} }.
\end{equation}
Furthermore, let $\mathbf{N_I}$ be the vector of $N^I_{ij}$ values
concatenated row-wise and let $\Sigma_I$ be the
variance-covariance matrix of $\mathbf{N_I}$ based on the correct
sampling distribution of the cell counts.
That is,
$\Sigma_I=\left(
\Cov\left[ N^I_{ij},N^I_{kl} \right] \right)$ where
$$\Cov\left[ N^I_{ij},N^I_{kl} \right]=
\frac{n}{\sqrt{n_i\,c_j\,n_k\,c_l}}\Cov\left[ N_{ij},N_{kl} \right]$$
with $\Cov\left[ N_{ij},N_{kl} \right]$ is as in Equation \eqref{eqn:VarNij} if
$(i,j)=(k,l)$ and as in Equation \eqref{eqn:cov-N11-N22} if
$(i,j)=(1,1)$ and $(k,l)=(2,2)$.
Since $\Sigma_I$ is not invertible,
I use its generalized inverse which is denoted by $\Sigma_I^-$ (\cite{searle:2006}).
Then the first version of segregation tests suggested by \cite{ceyhan:corrected} is
\begin{equation}
\label{eqn:X-square-I}
\X_I^2=\mathbf{N'_I}\Sigma_I^-\mathbf{N_I}
\end{equation}
which asymptotically has a $\chi^2_1$ distribution.

Under CSR independence, the expected values, variances,
and covariances related to $\X_I^2$ are as in the RL case,
except they are not only conditional on column sums (i.e., on $C_j=c_j$),
but also conditional on $Q$ and $R$.
Hence $\X_I^2$ has asymptotically $\chi^2_1$ distribution conditional on
column sums, $Q$ and $R$ under CSR independence.
Replacing $Q$ and $R$ by their empirical estimates given in Section \ref{sec:QandR},
I obtain the QR-adjusted version of this test which is denoted by $\X_{I,qr}^2$,
which is still conditional on column sums.

\subsection{Version II of the New Segregation Tests}
\label{sec:seg-test-II}
For cell $(i,j)$, let
\begin{equation}
\label{eqn:seg-test-II}
T^{II}_{ij}=N_{ij}-\frac{n_i\,n_j}{n} ~~~ \text{  and then let  }~~~
N^{II}_{ij}=\frac{T^{II}_{ij}}{\sqrt{n_i\,n_j/n}}=
\frac{\left( N_{ij}-n_i\,n_j/n \right)}{ \sqrt{n_i\,n_j/n}}.
\end{equation}
Furthermore, let $\mathbf{N_{II}}$ be the vector of
$N^{II}_{ij}$ concatenated row-wise and let
$\Sigma_{II}$ be the variance-covariance matrix of
$\mathbf{N_{II}}$ based on the correct sampling distribution of the cell counts.
That is, $\Sigma_{II}=\left(
\Cov\left[ N^{II}_{ij},N^{II}_{kl} \right] \right)$ where
$$\Cov\left[ N^{II}_{ij},N^{II}_{kl} \right]=
\frac{n}{\sqrt{n_i\,n_j\,n_k\,n_l}}\Cov\left[ N_{ij},N_{kl} \right].$$
Since $\Sigma_{II}$ is not invertible,
I use its generalized inverse $\Sigma_{II}^-$.
Then second version of the tests proposed by \cite{ceyhan:corrected} is
\begin{equation}
\label{eqn:X-square-II}
\X_{II}^2=\mathbf{N'_{II}}\Sigma_{II}^-\mathbf{N_{II}}
\end{equation}
which asymptotically has a $\chi^2_2$ distribution under RL.
Note that $\Sigma_{II}$ can be obtained from $\Sigma$
used in Equation \eqref{eqn:dix-chisq-2x2}
by multiplying $\Sigma$ entry-wise with the matrix
$C^{II}_M=\left( \frac{n}{\sqrt{n_i\,n_j\,n_k\,n_l}} \right)$.
This version of the segregation test is asymptotically equivalent
to Dixon's segregation test.

Under CSR independence, the expectations, variances, and
covariances related to $\X^2_{II}$ are as in the RL case,
but the variances and covariances are conditional on $Q$ and $R$.
Hence, the asymptotic distribution of $\X^2_{II}$
is also conditional on $Q$ and $R$.
Replacing $Q$ and $R$ with their empirical estimates,
I obtain the QR-adjusted version of this test
which is denoted by $\X^2_{II,qr}$ and is not conditional any more.

\subsection{Version III of the New Segregation Tests}
\label{sec:seg-test-III}
Notice that version I is a conditional test (conditional on column sums),
while version II is asymptotically equivalent to Dixon's test,
Furthermore, both Dixon's test and
version II incorporate only row sums (i.e., class sizes) in the NNCTs.

\cite{ceyhan:corrected} suggests another test statistic which uses both the
column sums (i.e., number of times a class serves as NN) and row sums and is not
conditional on the column sums.
Let
\begin{equation}
\label{eqn:seg-test-III}
T^{III}_{ij}=
\begin{cases}
N_{ij}-\frac{(n_i-1)}{(n-1)}\,C_j & \text{if $i=j$,}\\
N_{ij}-\frac{n_i}{(n-1)}\,C_j     & \text{if $i \not= j$.}
\end{cases}
\end{equation}
Let $\mathbf{N_{III}}$ be the vector of $T^{III}_{ij}$ values
concatenated row-wise and let $\Sigma_{III}$ be the
variance-covariance matrix of $\mathbf{N_{III}}$ based on the correct
sampling distribution of the cell counts.
That is,
$\Sigma_{III}=\left( \Cov\left[ T^{III}_{ij},T^{III}_{kl} \right] \right)$
where the explicit forms of $\Cov\left[ T^{III}_{ij},T^{III}_{kl} \right]$
are provided in (\cite{ceyhan:corrected}).
Since $\Sigma_{III}$ is not invertible,
I use its generalized inverse $\Sigma_{III}^-$.
Then the proposed test statistic by (\cite{ceyhan:corrected})
for overall segregation is the quadratic form
$\X_{III}^2=\mathbf{N'_{III}}\Sigma_{III}^-\mathbf{N_{III}}$
which asymptotically has a $\chi^2_1$ distribution.

Under CSR independence, the discussion related to and
derivation of $\X^2_{III}$ are as in the RL case;
however, the variance and covariance terms (hence the
asymptotic distribution) are conditional on $Q$ and $R$.
Replacing $Q$ and $R$ with their empirical estimates,
I obtain the QR-adjusted version of this test
which is denoted by $\X^2_{III,qr}$.

\begin{remark}
\label{rem:extension-to-multiple-classes}
\textbf{Extension to Multi-Class Case:}
So far, I have described the segregation tests for the two class case
in which the corresponding NNCT is of dimension $2\times 2$.
The cell counts for the diagonal cells have asymptotic normality.
For the off-diagonal cells, although the asymptotic normality is supported by Monte Carlo
simulation results (\cite{dixon:NNCTEco2002}),
it is not rigorously proven yet.
Nevertheless, if the asymptotic normality held for all $q^2$ cell counts in the NNCT,
under RL, Dixon's test and version II would have $\chi^2_{q(q-1)}$ distribution,
versions I and III would have $\chi^2_{(q-1)^2}$ distribution asymptotically.
Under CSR independence, these tests will have the corresponding asymptotic distributions
conditional on $Q$ and $R$.
The QR-adjusted versions can be obtained by replacing $Q$ and $R$
with their empirical estimates.
\end{remark}

\section{Empirical Significance Levels of NNCT-Tests under the CSR Independence}
\label{sec:emp-sign-level}
For the null case, $H_o:$ CSR independence, I simulate the CSR case only with
classes 1 and 2 (i.e., $X$ and $Y$) of sizes $n_1$ and $n_2$, respectively.
At each of $N_{mc}=10000$ replicates,
I generate data for some sample size combinations of $n_1,n_2 \in \{10,30,50,100\}$
points iid from $\U((0,1)\times (0,1))$.
These sample size combinations are chosen so that one can examine the
influence of small and large samples, and the relative abundance of the classes on the tests.
The corresponding test statistics are recorded at each Monte Carlo
replication for each sample size combination.
Then I record how many times the $p$-value is at or below $\al=.05$ for each test
to estimate the empirical size.
I present the empirical sizes for NNCT-tests in Table \ref{tab:MC-emp-sig-level-NNCT},
where $\widehat{\al}_D$ is the empirical significance level
for Dixon's test, $\widehat{\al}_I,\,\widehat{\al}_{II}$ and $\widehat{\al}_{III}$
are for versions I, II, and III, respectively,
and $\widehat{\al}_{D,qr}$, $\widehat{\al}_{I,qr},\,\widehat{\al}_{II,qr}$ and $\widehat{\al}_{III,qr}$
are for the corresponding QR-adjusted versions.
The empirical sizes significantly smaller (larger) than .05 are
marked with $^c$ ($^{\ell}$),
which indicate that the corresponding test is conservative (liberal).
The asymptotic normal approximation
to proportions is used in determining the significance of the
deviations of the empirical size estimates from the nominal level of .05.
For these proportion  tests, I also use $\alpha=.05$ to test against
empirical size being equal to .05.
With $N_{mc}=10000$, empirical
sizes less (greater) than .0464 (.0536) are deemed conservative (liberal) at $\alpha=.05$ level.

Observe that the (unadjusted) NNCT-tests are about the desired level
(or size) when $n_1$ and $n_2$ are both $\ge 30$,
and mostly conservative otherwise.
The same trend holds for the QR-adjusted versions.
Furthermore, comparing the empirical sizes of QR-adjusted versions
with those of unadjusted ones,
I see that for almost all cases they are not significantly different
(at $\al=.05$ based on tests on equality of the proportions).

\begin{table}[ht]
\centering
\begin{tabular}{|c||c|c|c|c||c|c|c|c|}
\hline
\multicolumn{9}{|c|}{Empirical significance levels of the NNCT-tests} \\
\hline
&\multicolumn{4}{|c||}{conditional (i.e., unadjusted)}&\multicolumn{4}{|c|}{unconditional (i.e., QR-adjusted)} \\
\hline
$(n_1,n_2)$ &  $\widehat{\al}_D$ & $\widehat{\al}_I$ & $\widehat{\al}_{II}$ & $\widehat{\al}_{III}$
 &  $\widehat{\al}_{D,qr}$ & $\widehat{\al}_{I,qr}$ & $\widehat{\al}_{II,qr}$ & $\widehat{\al}_{III,qr}$ \\
\hline
 (10,10)   & .0432$^c$ & .0593$^{\ell}$ & .0461$^c$ & .0439$^c$  & .0470 & .0595$^{\ell}$ & .0486 & .0365$^{c,<}$ \\
\hline
 (10,30)   & .0440$^c$ & .0451$^c$ & .0421$^c$ & .0410$^c$  & .0411$^c$ & .0465 & .0381$^c$ & .0461$^{c,>}$ \\
\hline
 (10,50)   & .0482 & .0335$^c$ & .0423$^c$ & .0397$^c$  & .0497 & .0345$^c$ & .0411$^c$ & .0431$^c$ \\
\hline
 (30,10)   &  .0390$^c$ &  .0411$^c$ &  .0383$^c$ &  .0391$^c$  & .0402$^c$ & .0423$^c$ & .0379$^c$ & .0436$^c$ \\
\hline
 (30,30)   & .0464 & .0544$^{\ell}$ & .0476 & .0427$^c$  & .0492 & .0552$^{\ell}$ & .0478 & .0409$^c$ \\
\hline
 (30,50)   & .0454$^c$ & .0507 & .0481 & .0504  & .0411$^c$ & .0517 & .0464 & .0515 \\
\hline
 (50,10)   &  .0529 &  .0326$^c$ &  .0468 &  .0379$^c$  & .0510 & .0334$^c$ & .0428$^c$ & .0402$^c$ \\
\hline
 (50,30)   &  .0429$^c$ &  .0494 &  .0468 &  .0469 & .0405$^c$ & .0518 & .0466 & .0492 \\
\hline
 (50,50)   & .0508 & .0494 & .0497 & .0499 & .0528 & .0494 & .0524 & .0488 \\
\hline
 (50,100)  &  .0560$^{\ell}$ &  .0501 &  .0564$^{\ell}$ &  .0516 & .0556$^{\ell}$ & .0493 & .0573 & .0494 \\
\hline
 (100,50)  &  .0483 &  .0463$^c$ &  .0492 &  .0479 & .0495 & .0457 & .0501 & .0460 \\
\hline
 (100,100) &  .0504 &  .0524 &  .0519 &  .0489 & .0513 & .0524 & .0523 & .0463$^c$ \\
\hline
\end{tabular}
\caption{
\label{tab:MC-emp-sig-level-NNCT}
The empirical significance levels for Dixon's, and the new versions of the NNCT-tests
by (\cite{ceyhan:corrected})
as well as their QR-adjusted versions
based on 10000 Monte Carlo simulations of CSR independence pattern.
$\widehat{\al}_D$ stands for the empirical significance level
for Dixon's test, $\widehat{\al}_I,\,\widehat{\al}_{II}$ and $\widehat{\al}_{III}$
for versions I, II, and III, respectively;
and $\widehat{\al}_{D,qr}$, $\widehat{\al}_{I,qr},\,\widehat{\al}_{II,qr}$ and $\widehat{\al}_{III,qr}$
stand for the corresponding QR-adjusted versions.
($^c$ ($^{\ell}$): the empirical size is significantly smaller (larger) than .05;
i.e., the test is conservative (liberal).
$^<$ ($^>$): the empirical size of QR-adjusted version is
significantly smaller (larger) than that of unadjusted version.)
}
\end{table}

\section{Empirical Power Analysis}
\label{sec:emp-power}
To evaluate the power performance of the QR-adjusted and unadjusted NNCT-tests,
I only consider alternatives against the CSR pattern.
That is, the points are generated in such a way that
they are from an inhomogeneous Poisson process in a region of interest (unit square
in the simulations) for at least one class.
Furthermore, the tests considered in this article seem to
have the desired nominal level for large samples under CSR,
and QR-adjustment is not necessary under the RL pattern.
Hence I avoid the alternatives against the RL pattern;
i.e., I do not consider non-random labeling of a fixed set of points
that would result in segregation or association.

\subsection{Empirical Power Analysis under Segregation Alternatives}
\label{sec:emp-power-seg}
For the segregation alternatives (against the CSR pattern), three cases are considered.
I generate $X_i \stackrel{iid}{\sim} \U((0,1-s)\times(0,1-s))$ for $i=1,2,\ldots,n_1$
and $Y_j \stackrel{iid}{\sim} \U((s,1)\times(s,1))$ for $j=1,2,\ldots,n_2$.
In the pattern generated, appropriate choices of
$s$ will imply $X_i$ and $Y_j$ to be more segregated than expected under CSR.
That is, it will be  more likely to have $(X,X)$
NN pairs than mixed NN pairs (i.e., $(X,Y)$ or $(Y,X)$ pairs).
The three values of $s$ I consider constitute
the three segregation alternatives:
\begin{equation}
\label{eqn:assoc-alt}
H_S^{I}: s=1/6,\;\;\; H_S^{II}: s=1/4, \text{ and } H_S^{III}: s=1/3.
\end{equation}
Observe that, from $H_S^I$ to $H_S^{III}$ (i.e., as $s$ increases), the segregation gets stronger
in the sense that $X$ and $Y$ points tend to form one-class clumps or clusters.
By construction, the points are uniformly generated, hence exhibit homogeneity
with respect to their supports for each class,
but with respect to the unit square these alternative patterns are examples of
departures from first-order homogeneity which implies
segregation of the classes $X$ and $Y$.
The simulated segregation patterns are symmetric in the sense that,
$X$ and $Y$ classes are equally segregated (or clustered) from each other.

\begin{figure}[ht]
\centering
\rotatebox{-90}{ \resizebox{2.1 in}{!}{\includegraphics{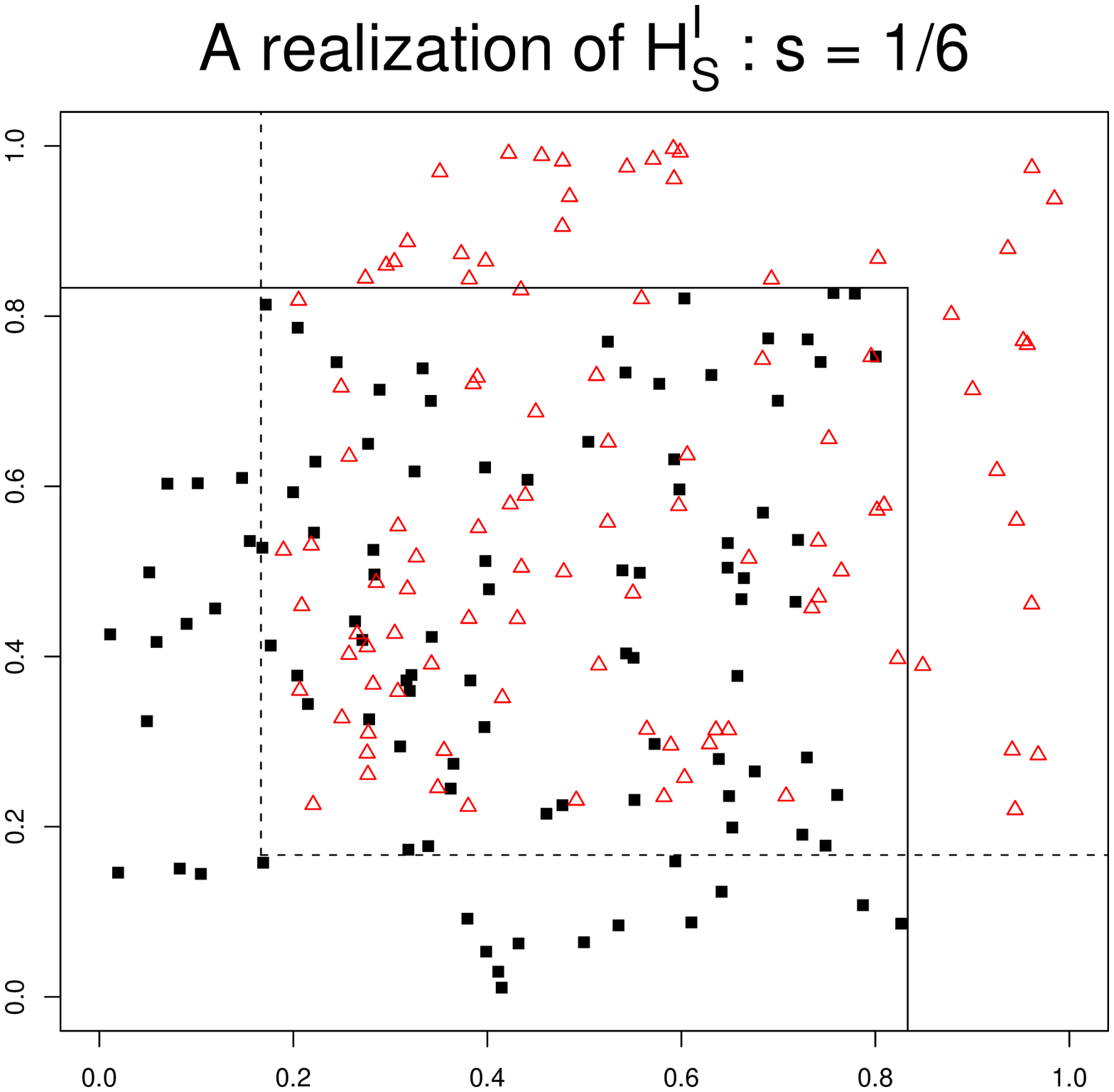} }}
\rotatebox{-90}{ \resizebox{2.1 in}{!}{\includegraphics{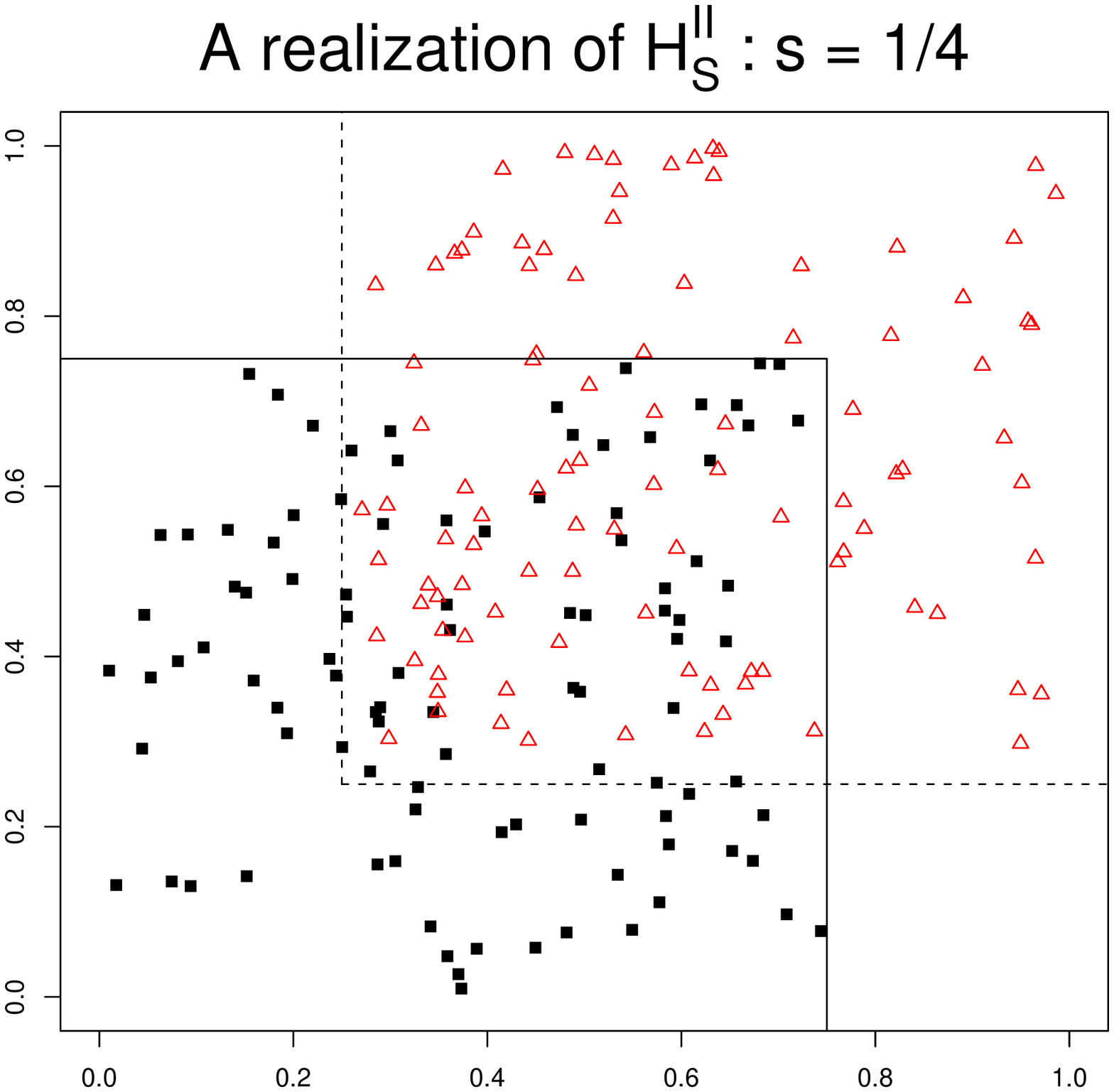} }}
\rotatebox{-90}{ \resizebox{2.1 in}{!}{\includegraphics{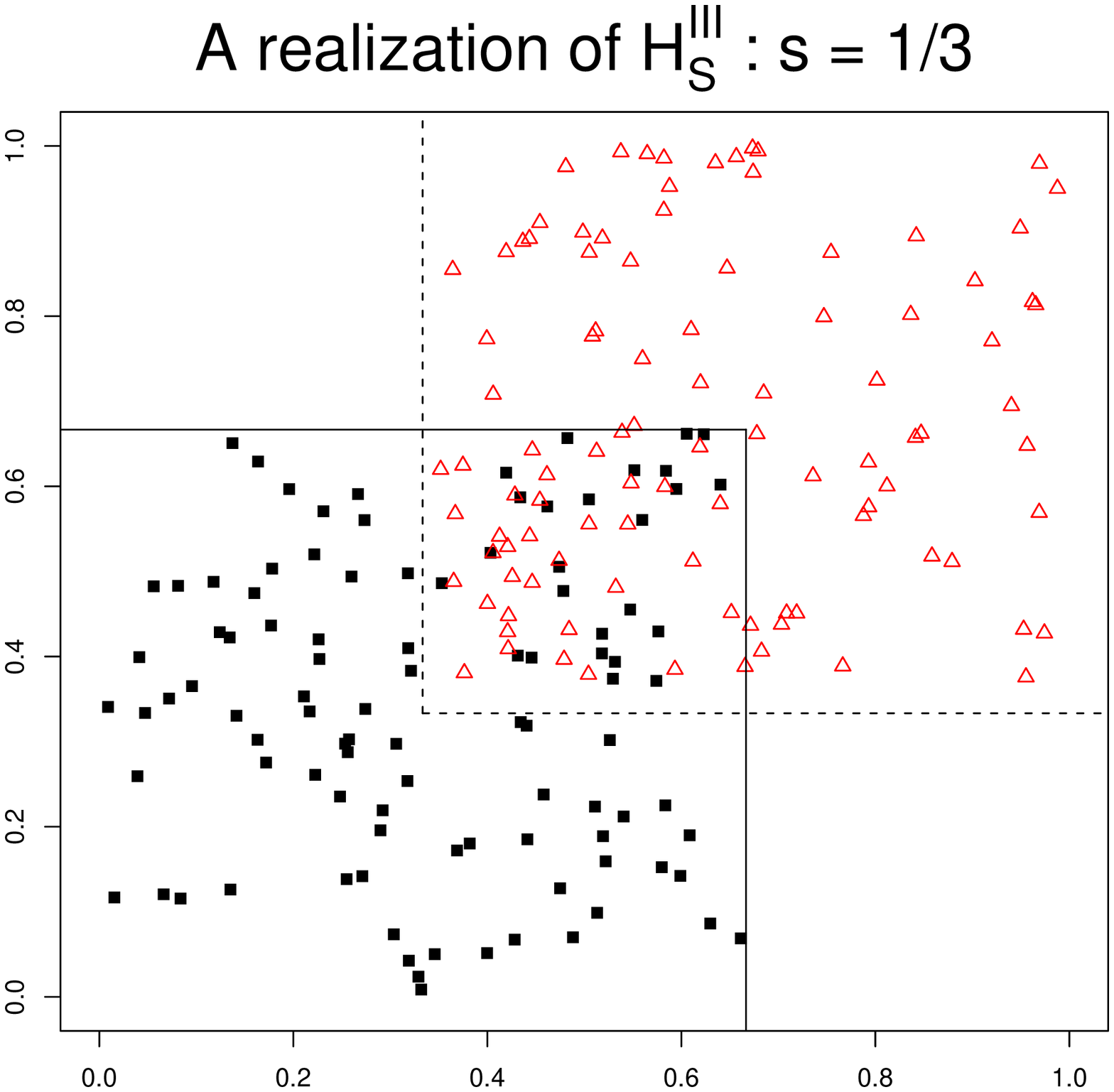} }}
 \caption{
\label{fig:SegAlt}
Three realizations for $H_S^{I}: s=1/6$, $H_S^{II}: s=1/4$,
and $H_S^{III}: s=1/3$ with $n_1=100$ $X$ points (solid squares $\blacksquare$)
and $n_2=100$ $Y$ points (triangles $\triangle$).}
\end{figure}

\begin{figure}[]
\centering
\rotatebox{-90}{ \resizebox{2.1 in}{!}{\includegraphics{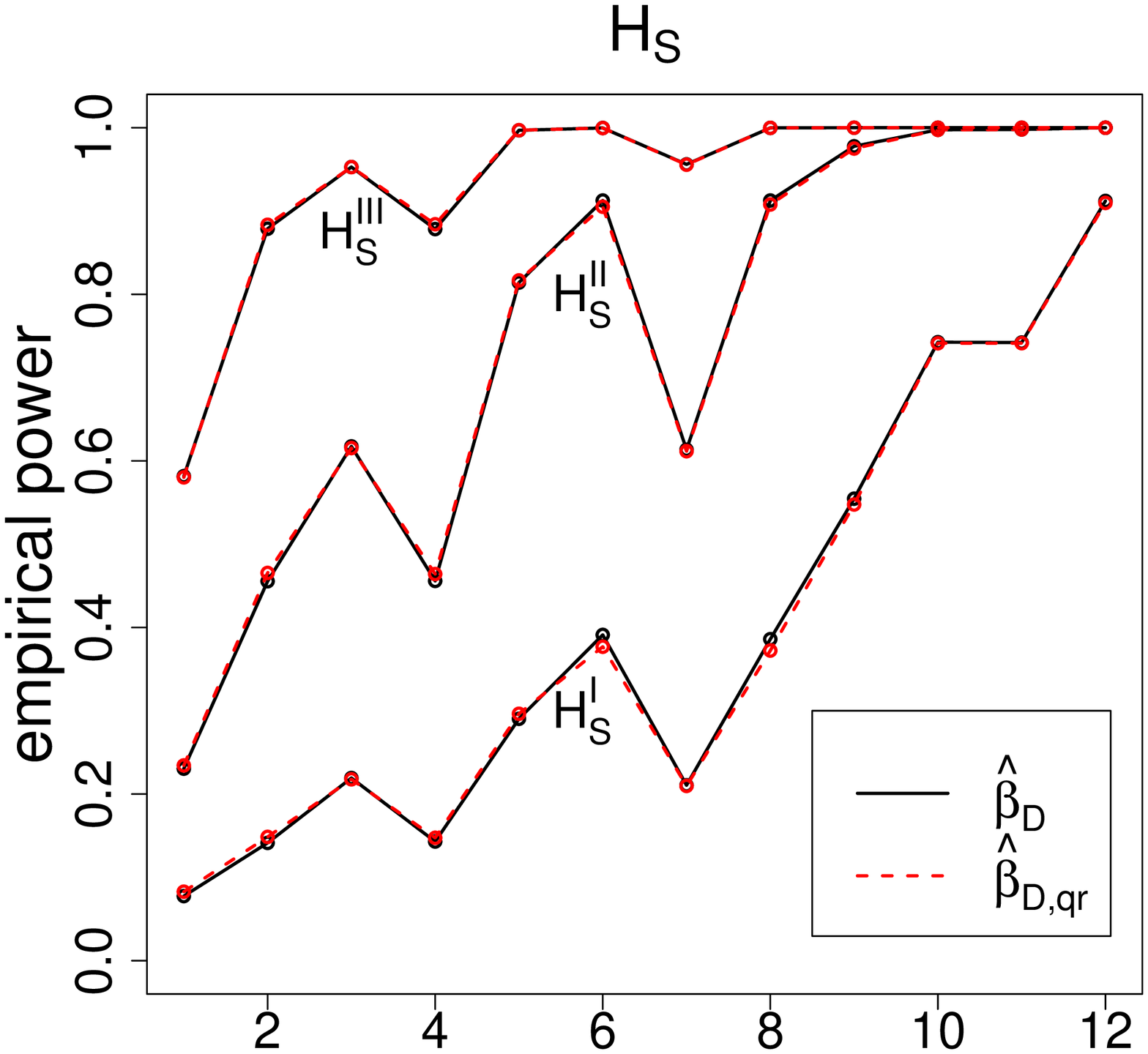} }}
\rotatebox{-90}{ \resizebox{2.1 in}{!}{\includegraphics{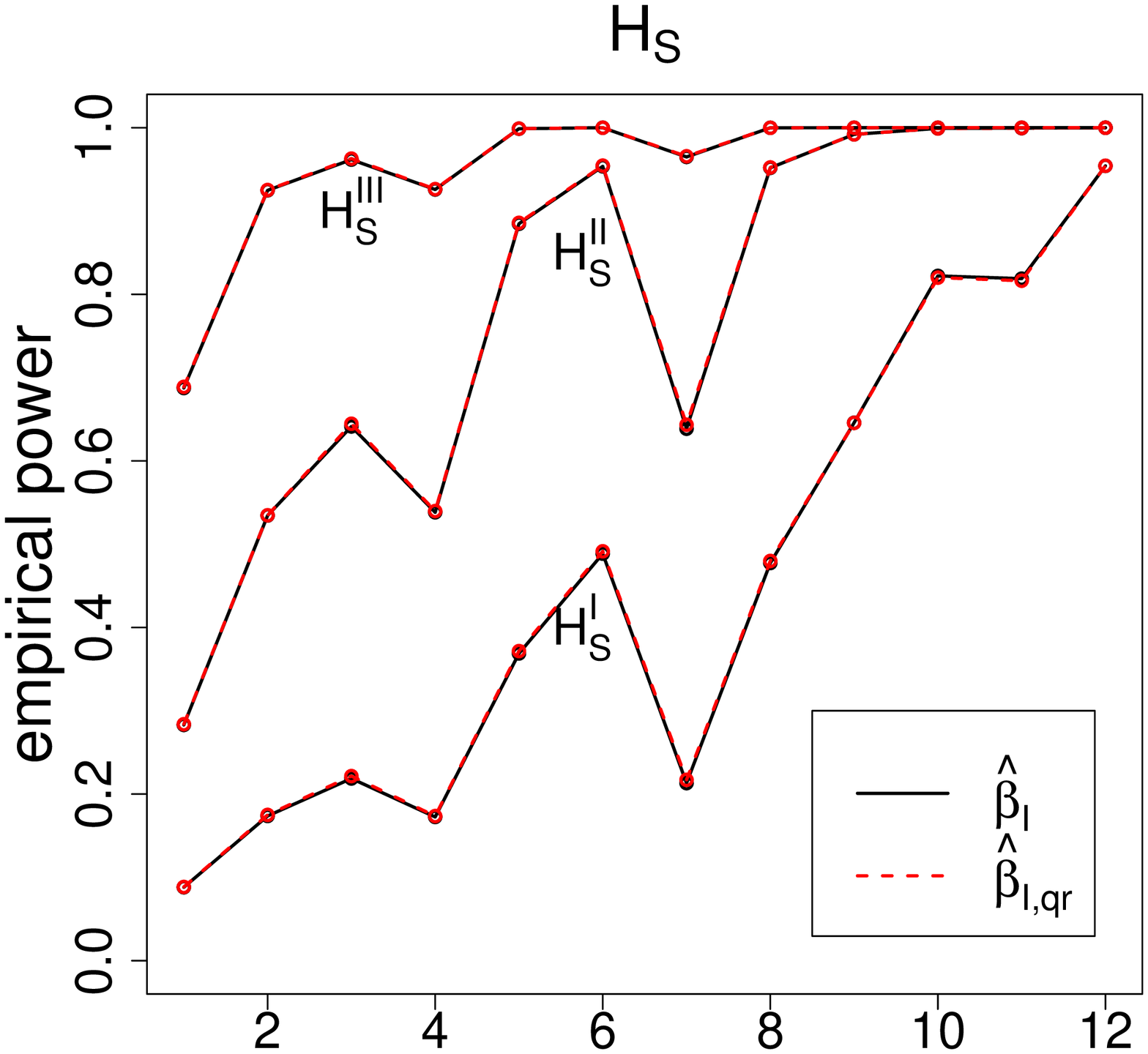} }}
\rotatebox{-90}{ \resizebox{2.1 in}{!}{\includegraphics{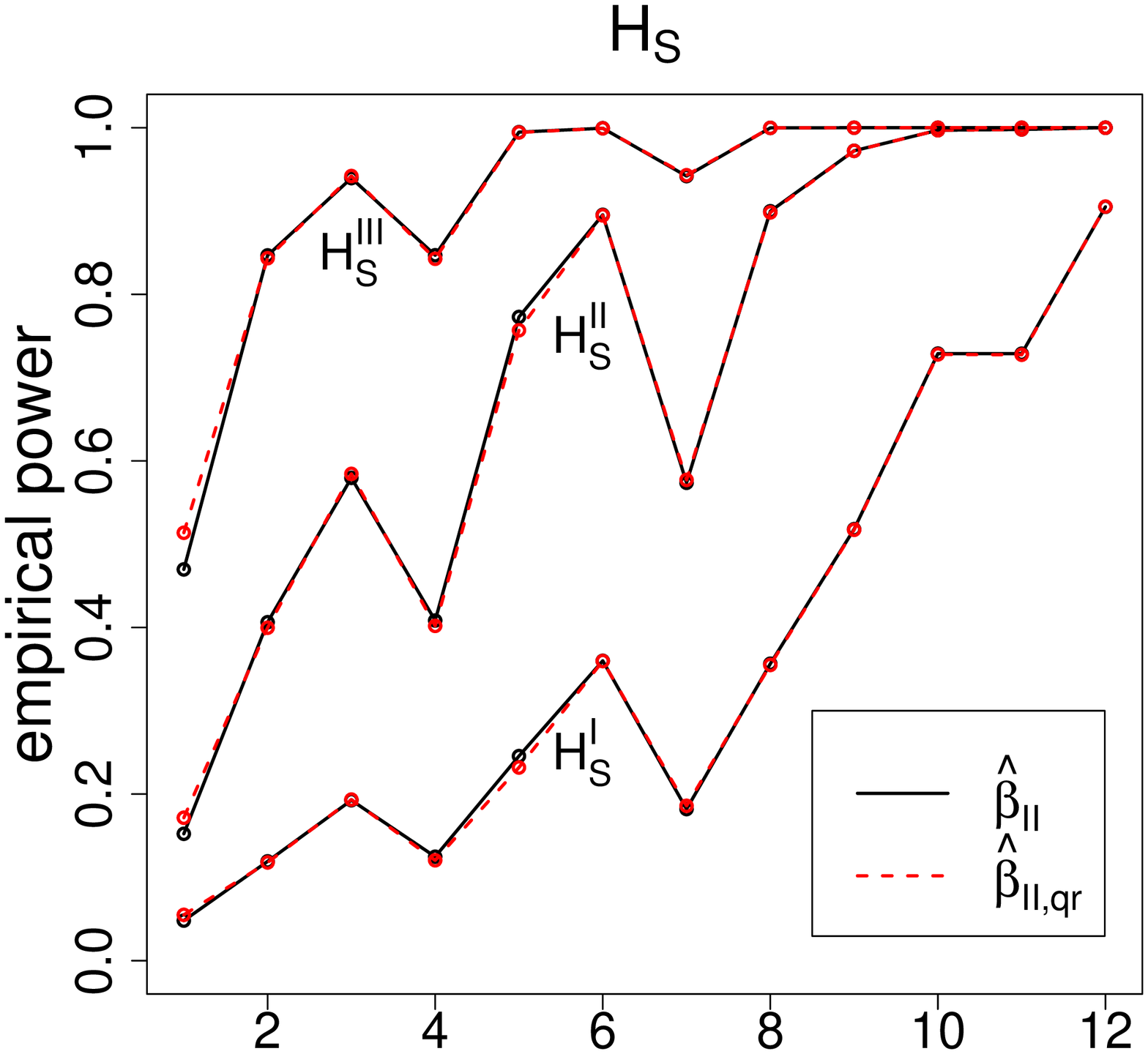} }}
\rotatebox{-90}{ \resizebox{2.1 in}{!}{\includegraphics{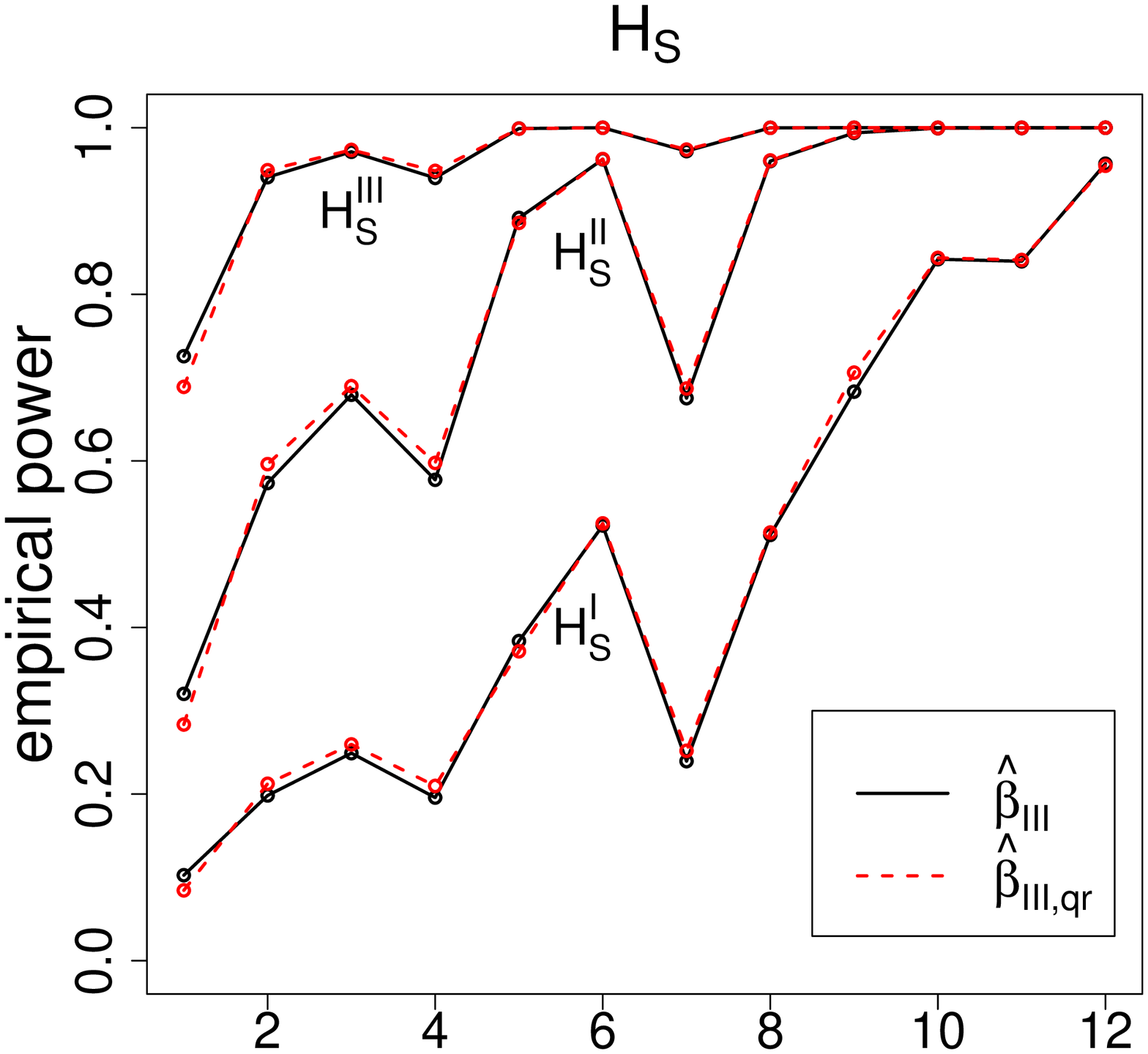} }}
\caption{
\label{fig:Power-Est-Seg}
Empirical power estimates for the QR-adjusted and unadjusted NNCT-tests
based on 10000 Monte Carlo replications under the segregation alternatives.
The numbers in the horizontal axis labels represent sample (i.e., class) size combinations:
1=(10,10), 2=(10,30), 3=(10,50), 4=(30,10), 5=(30,30), 6=(30,50),
7=(50,10), 8=(50,30), 9=(50,50), 10=(50,100), 11=(100,50), 12=(100,100).}
\end{figure}

The power estimates against the sample size combinations
are presented in Figure \ref{fig:Power-Est-Seg},
where $\bh_D$ is for Dixon's test, $\bh_I$, $\bh_{II}$, and $\bh_{III}$
are for versions I, II , and III, respectively,
and the QR-adjusted versions are indicated by $qr$ in their subscripts.
Observe that, as $n=(n_1+n_2)$ gets larger,
the power estimates get larger.
For the same $n=(n_1+n_2)$ values, the power estimate is larger
for classes with similar sample sizes.
Furthermore, as the segregation gets stronger,
the power estimates get larger.
The NNCT-tests have about the same power performance under these segregation alternatives.
Notice also that for small samples the power estimates of the QR-adjusted versions are slightly larger
but for other sample size combinations the power estimates for the QR-adjusted versions
and the unadjusted versions are virtually indistinguishable.

\subsection{Empirical Power Analysis under Association Alternatives}
\label{sec:emp-power-assoc}
For the association alternatives (against the CSR pattern), I also consider three cases.
First, I generate $X_i \stackrel{iid}{\sim} \U((0,1)\times(0,1))$ for $i=1,2,\ldots,n_1$.
Then I generate $Y_j$ for $j=1,2,\ldots,n_2$ as follows.
For each $j$, I pick an $i$ randomly, then generate $Y_j$ as
$X_i+R_j\,(\cos T_j, \sin T_j)'$ where
$R_j \stackrel{iid}{\sim} \U(0,r)$ with $r \in (0,1)$ and
$T_j \stackrel{iid}{\sim} \U(0,2\,\pi)$.
In the pattern generated, appropriate choices of
$r$ will imply $Y_j$ and $X_i$ are more associated than expected.
That is, it will be  more likely to have $(X,Y)$
NN pairs than self NN pairs (i.e., $(X,X)$ or $(Y,Y)$).
The three values of $r$ I consider constitute
the three association alternatives:
\begin{equation}
\label{eqn:assoc-alt}
H_A^{I}: r=1/4,\;\;\; H_A^{II}: r=1/7, \text{ and } H_A^{III}: r=1/10.
\end{equation}
Observe that, from $H_A^I$ to $H_A^{III}$ (i.e., as $r$ decreases),
the association gets stronger
in the sense that $X$ and $Y$ points tend to occur
together more and more frequently.
By construction,
$X$ points are from a homogeneous Poisson process with respect to the unit square,
while $Y$ points exhibit inhomogeneity in the same region.
Furthermore, these alternative patterns are examples of
departures from second-order homogeneity which implies
association of the class $Y$ with class $X$.

\begin{figure}[ht]
\centering
\rotatebox{-90}{ \resizebox{2.1 in}{!}{\includegraphics{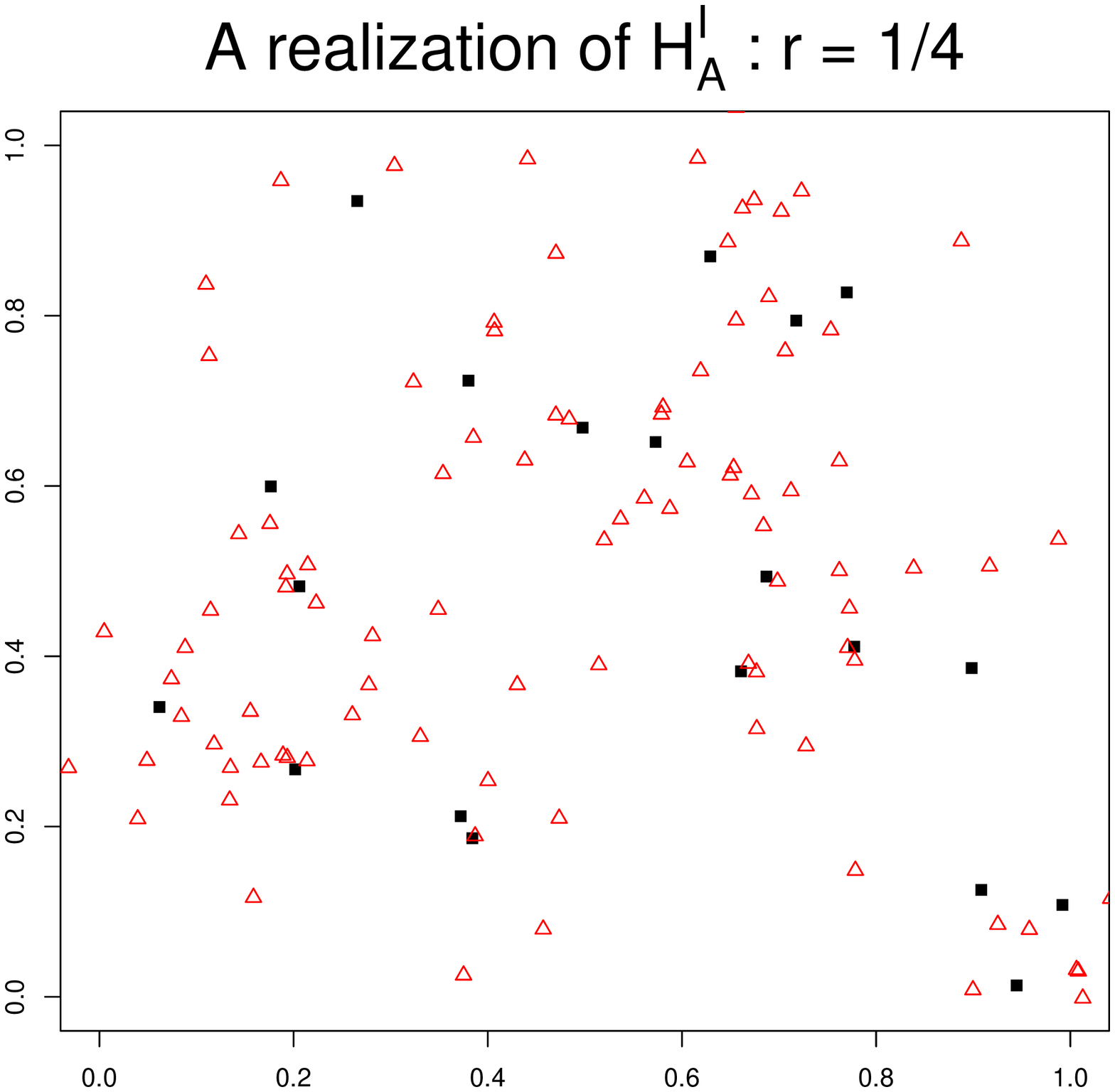} }}
\rotatebox{-90}{ \resizebox{2.1 in}{!}{\includegraphics{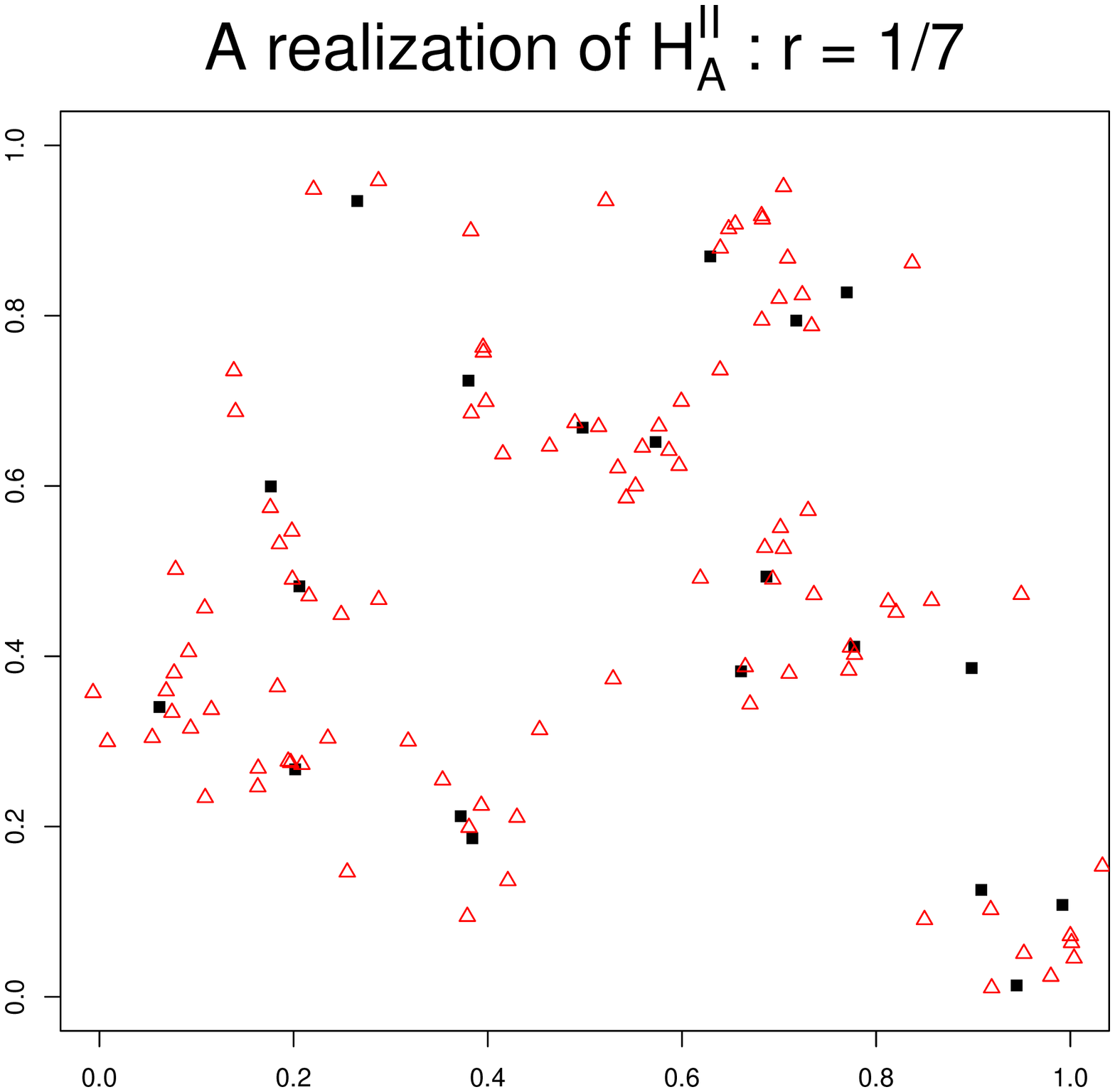} }}
\rotatebox{-90}{ \resizebox{2.1 in}{!}{\includegraphics{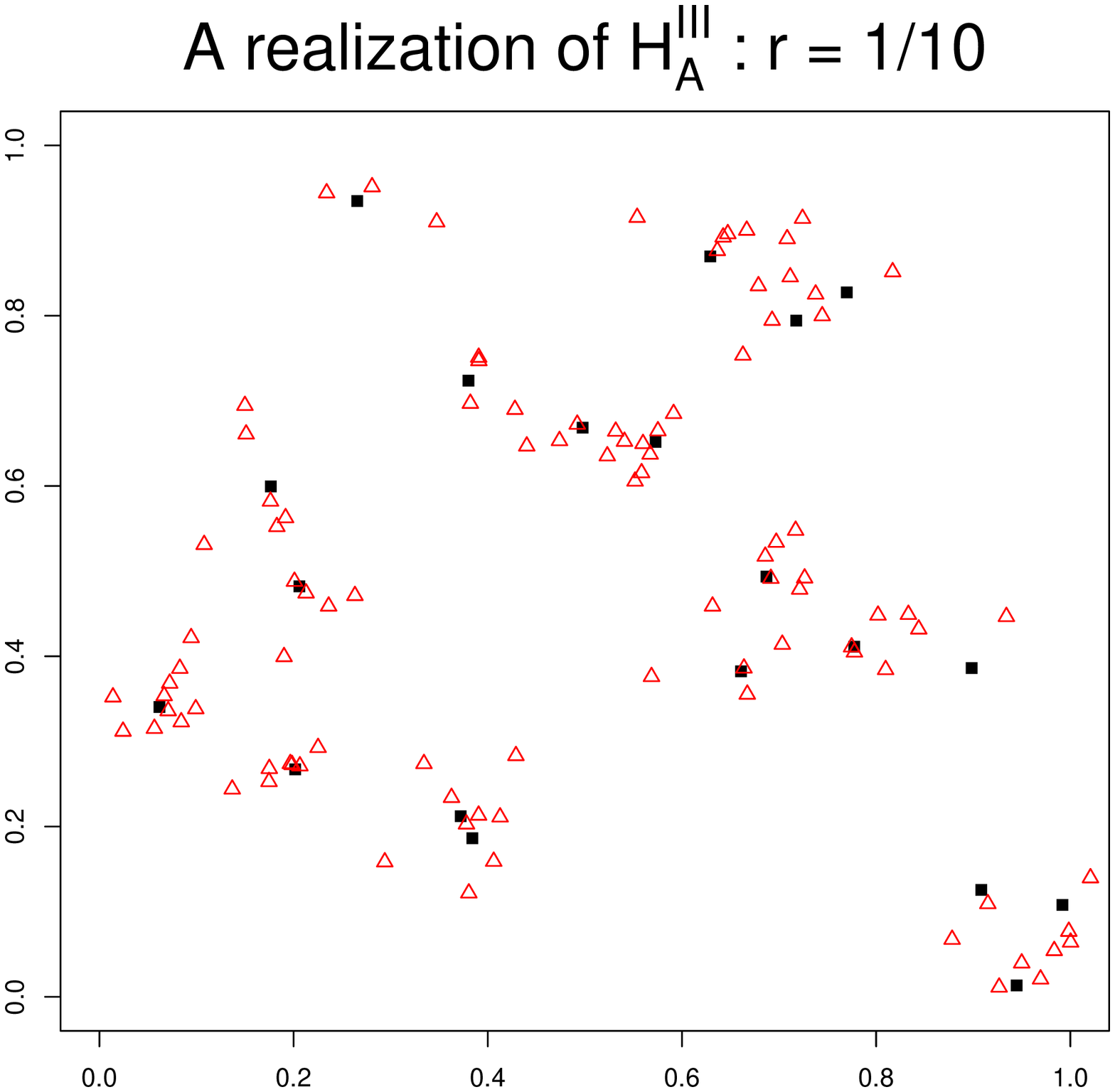} }}
 \caption{
\label{fig:AssocAlt}
Three realizations for $H_A^{I}: r=1/4$, $H_A^{II}: s=1/7$,
and $H_A^{III}: r=1/10$ with $n_1=20$ $X$ points (solid squares $\blacksquare$)
and $n_2=100$ $Y$ points (triangles $\triangle$).}
\end{figure}

\begin{figure}[]
\centering
\rotatebox{-90}{ \resizebox{2.1 in}{!}{\includegraphics{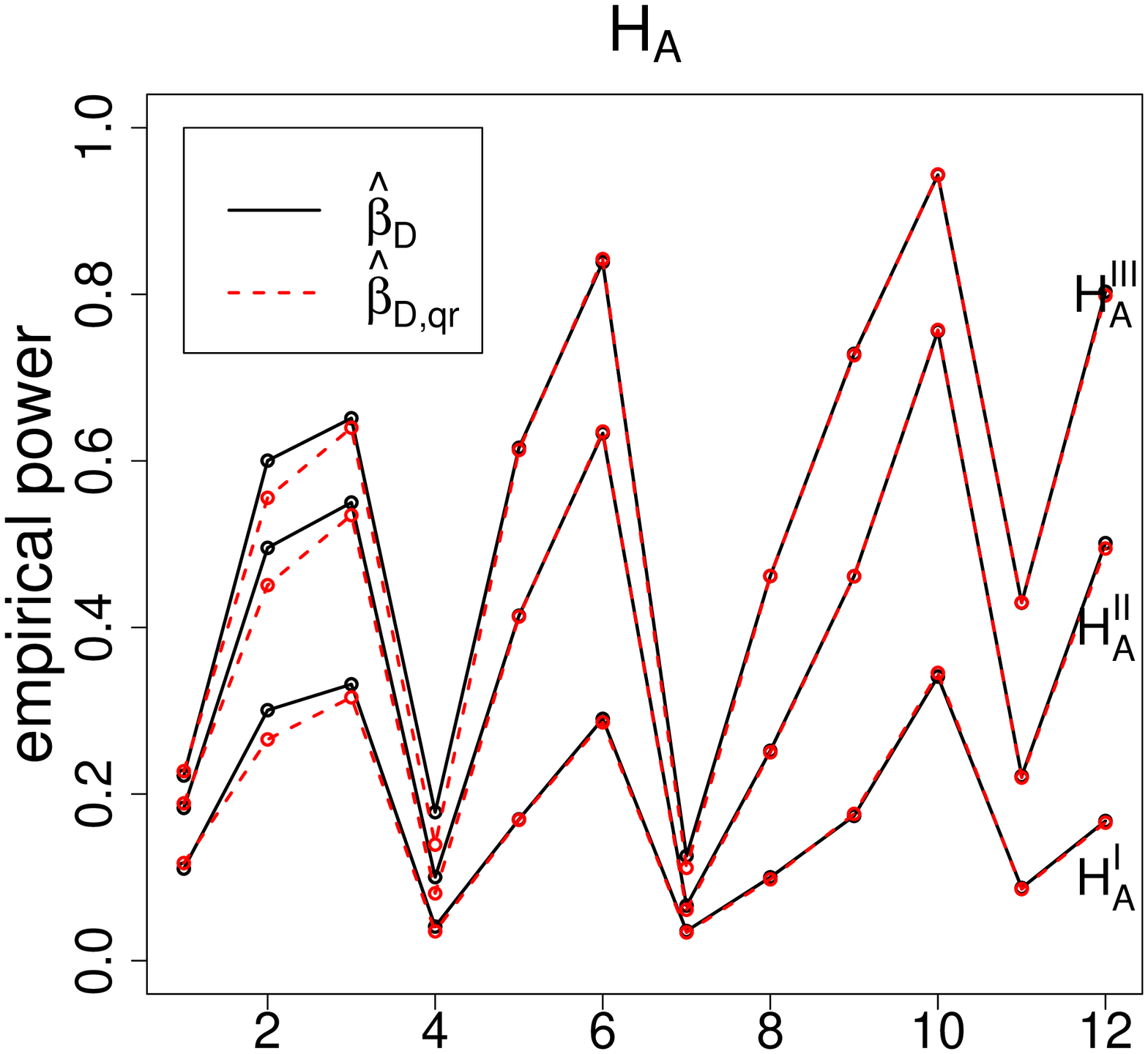} }}
\rotatebox{-90}{ \resizebox{2.1 in}{!}{\includegraphics{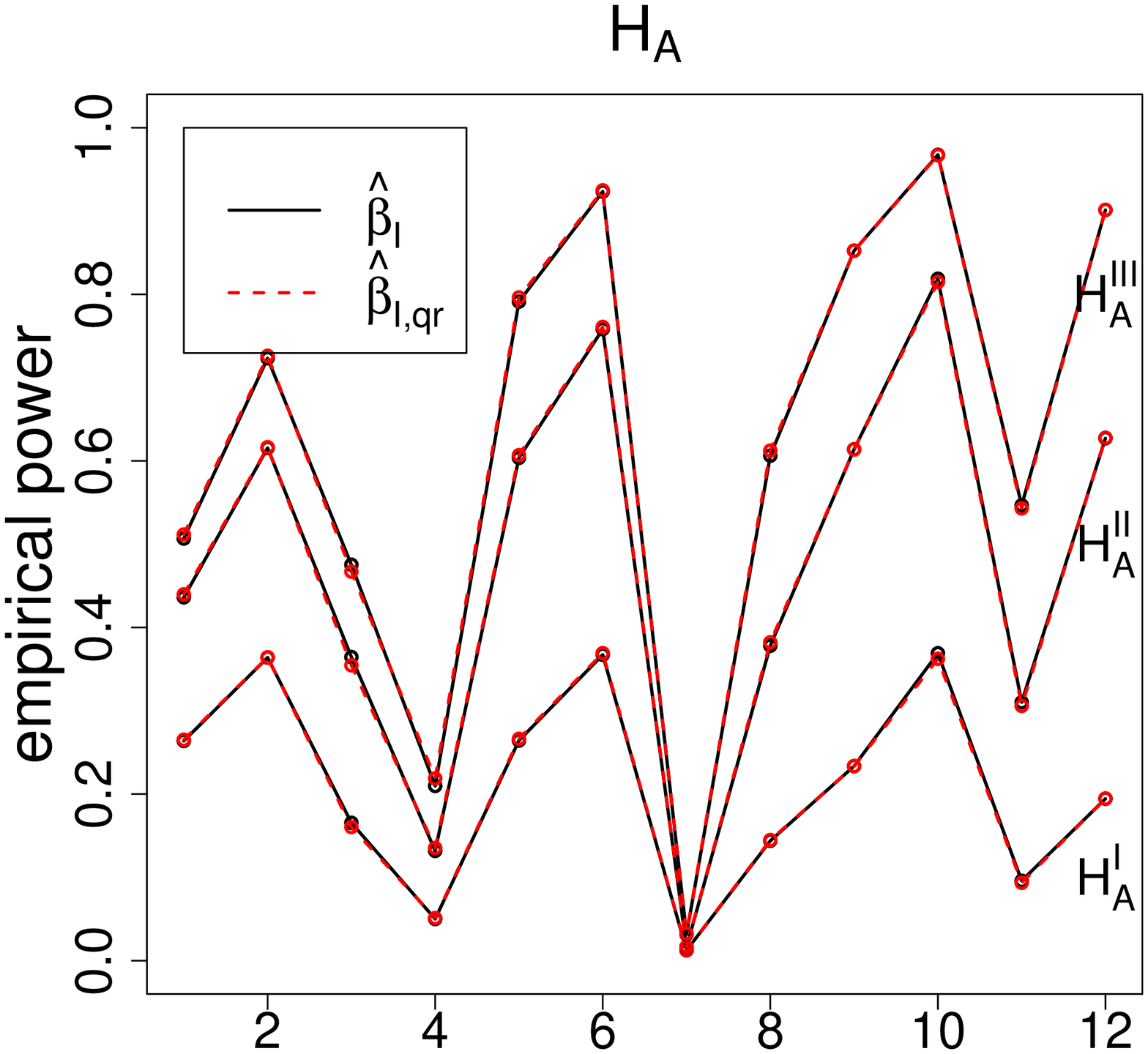} }}
\rotatebox{-90}{ \resizebox{2.1 in}{!}{\includegraphics{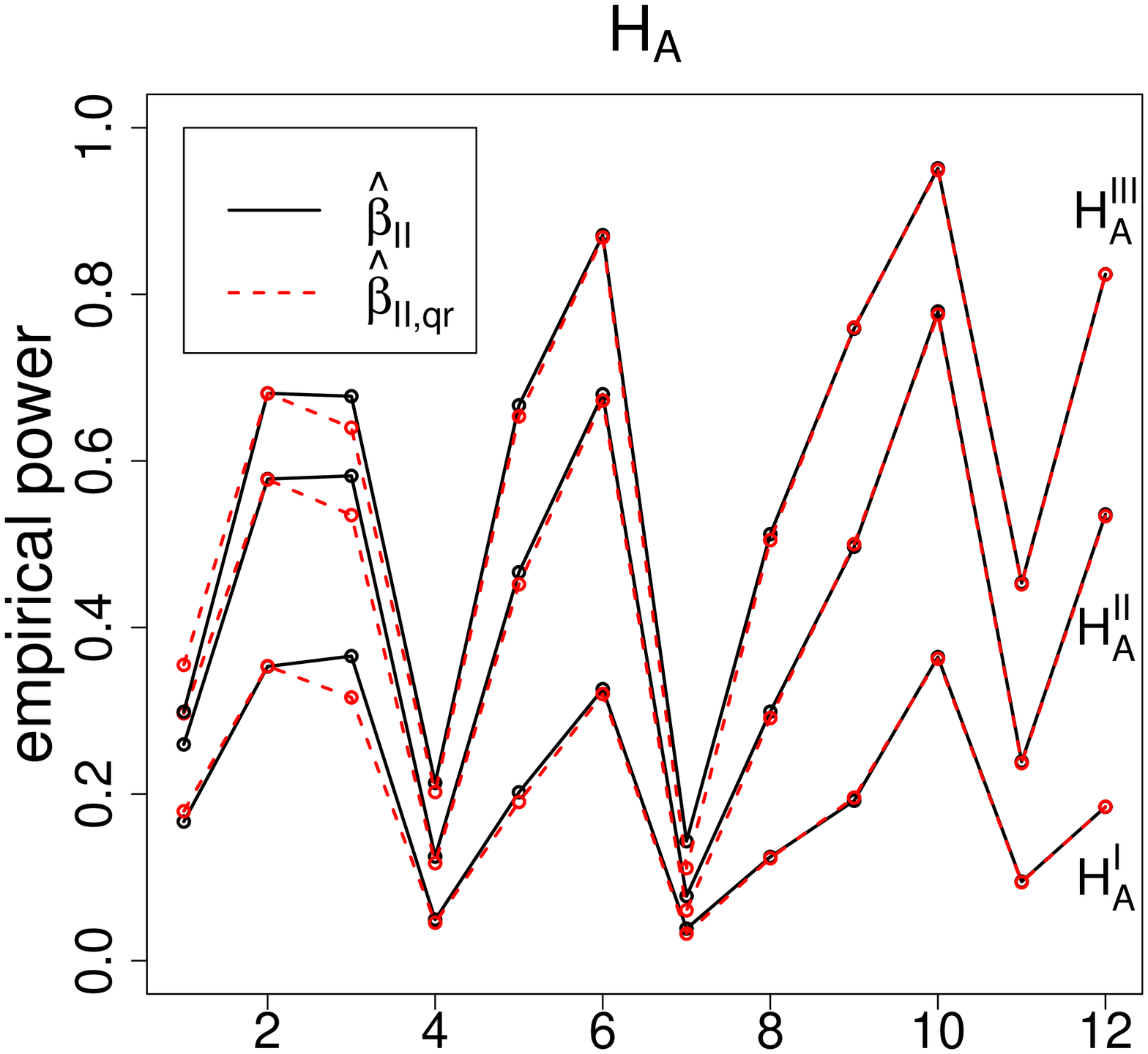} }}
\rotatebox{-90}{ \resizebox{2.1 in}{!}{\includegraphics{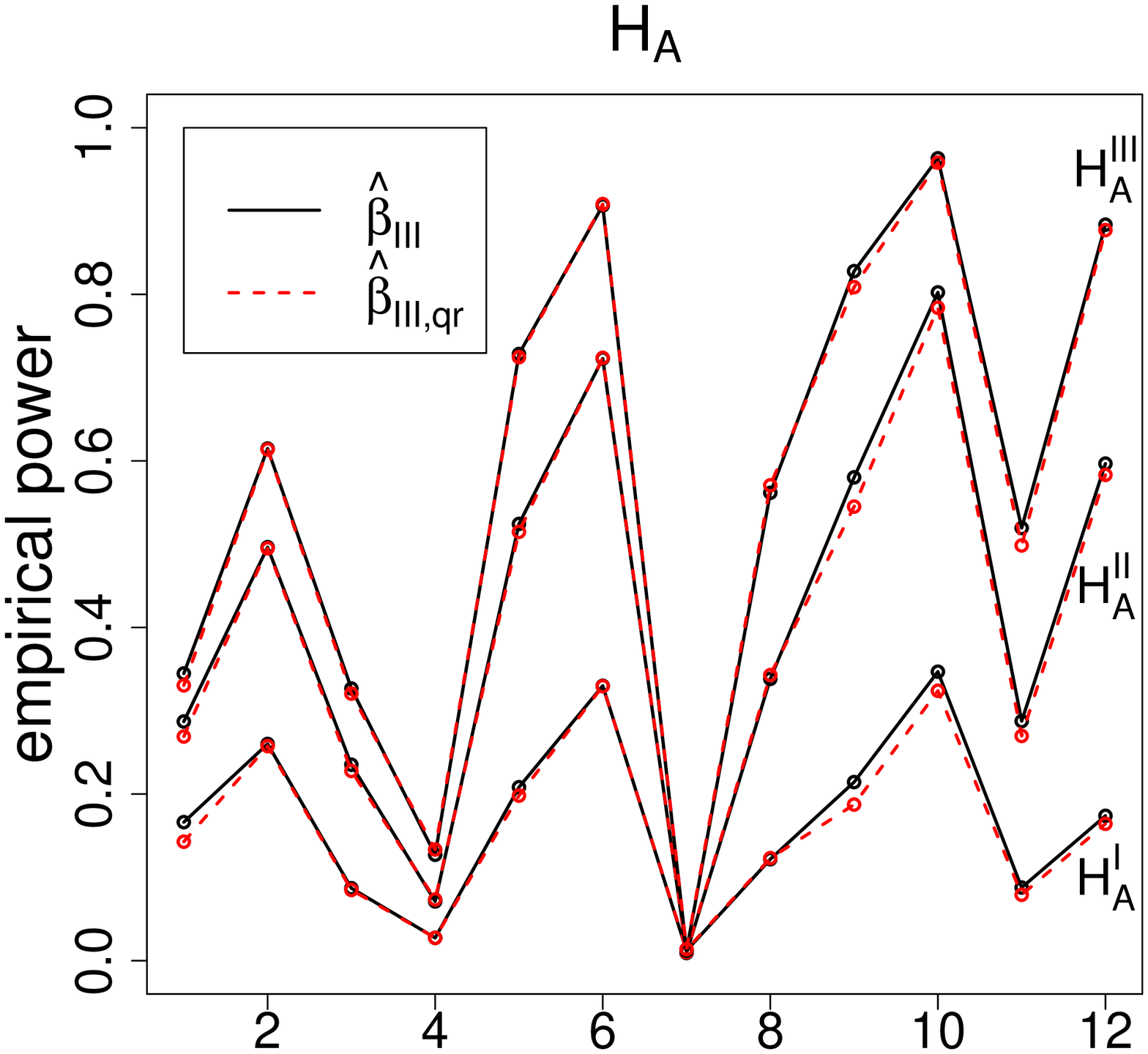} }}
\caption{
\label{fig:Power-Est-Assoc}
Empirical power estimates for the QR-adjusted and unadjusted NNCT-tests
under the association alternatives.
The numbers in the horizontal axis labels represent sample (i.e., class) size combinations:
1=(10,10), 2=(10,30), 3=(10,50), 4=(30,10), 5=(30,30), 6=(30,50),
7=(50,10), 8=(50,30), 9=(50,50), 10=(50,100), 11=(100,50), 12=(100,100).}
\end{figure}

The power estimates under the association alternatives
are presented in Figure \ref{fig:Power-Est-Assoc},
where labeling is as in Figure \ref{fig:Power-Est-Seg}.
Observe that, for similar sample sizes as $n=(n_1+n_2)$ gets larger,
the power estimates get larger at each association alternative.
Furthermore, as the association gets stronger,
the power estimates get larger at each sample size combination.
The NNCT-tests have about the same power estimates under these
association alternatives.
Furthermore the QR-adjusted versions of the tests virtually
have the same power estimates as the unadjusted versions;
for the smaller samples QR-adjusted version has slightly lower power estimates.

\begin{remark}
\label{rem:MC-emp-size-CSR}
\textbf{Main Result of Monte Carlo Simulation Analysis:}
Based on the simulation results under CSR independence of the points,
I observe that none of the NNCT-tests I consider
has the desired level when at least one sample size is small so that
the cell count(s) in the corresponding NNCT have a high probability of being $\le 5$.
This usually corresponds to the case that at least
one sample size is $\leq 10$ or the sample sizes are very different
in the simulation study.
When sample sizes are small (hence the corresponding cell counts are $\leq 5$),
the asymptotic approximation of the NNCT-tests is not appropriate.
So \cite{dixon:1994} recommends Monte Carlo randomization
for his test when some cell count(s) are $\le 5$ in a NNCT.
I extend this recommendation for all the NNCT-tests discussed in this article.
Furthermore, among the NNCT-tests, Dixon's and version III tests seem to be affected
by the QR-adjustment more than the other tests in terms of empirical size.
But QR-adjustment does not necessarily improve
the results of the NNCT-analysis under CSR independence,
as the empirical sizes of the adjusted and unadjusted versions are not significantly different.
Furthermore, the QR-adjustment does not significantly improve the power performance
under segregation and association alternatives.
In fact the power estimates of QR-adjusted and unadjusted tests were about the same under these alternatives.
\end{remark}

\section{Examples}
\label{sec:examples}
I illustrate the tests on two examples:
an ecological data set, namely swamp tree data (\cite{good:1982}),
and an artificial data set.

\subsection{Swamp Tree Data}
\label{sec:swamp-data}

\begin{figure}[ht]
\centering
\rotatebox{-90}{ \resizebox{2.5 in}{!}{\includegraphics{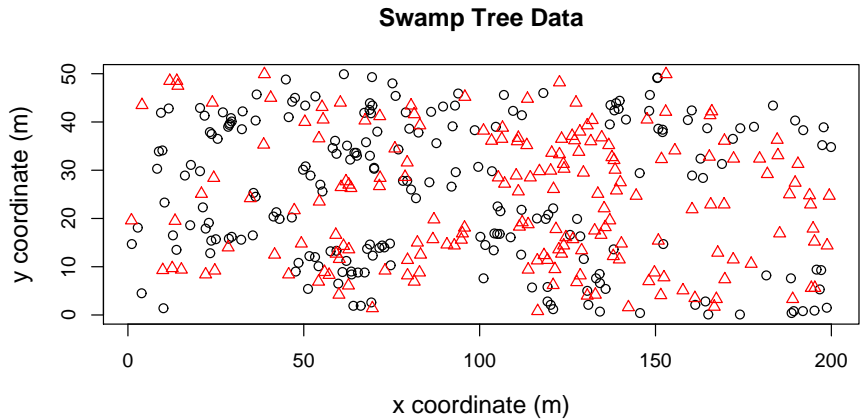} }}
\caption{
\label{fig:SwampTrees}
The scatter plot of the locations of water tupelos (circles $\circ$)
and black gums (triangles $\triangle$).}
\end{figure}

\cite{good:1982} considered the spatial patterns of tree species
along the Savannah River, South Carolina, U.S.A.
From this data, \cite{dixon:NNCTEco2002} used a single 50m $\times$ 200m rectangular plot
to illustrate his tests.
All live or dead trees with 4.5 cm or more dbh (diameter at breast height)
were recorded together with their species.
Hence it is an example of a realization of a marked multi-variate point pattern.
The plot contains 13 different tree species,
four of which comprise over 90 \% of the 734 tree stems.
The remaining tree stems were categorized as ``other trees".
The plot consists of 215 water tupelo (\emph{Nyssa aquatica}),
205 black gum (\emph{Nyssa sylvatica}), 156 Carolina ash (\emph{Fraxinus caroliniana}),
98 bald cypress (\emph{Taxodium distichum}), and 60 stems of 8 additional species (i.e., other species).
I will only consider live trees from the two most frequent tree species in this data set
(i.e., water tupelos and black gums).
So a $2 \times 2$ NNCT-analysis is conducted for this data set.
If segregation among the less frequent species were important,
a more detailed $5 \times 5$ or a $12 \times 12$ NNCT-analysis should be performed.
The locations of these trees in the study region are plotted in Figure \ref{fig:SwampTrees}
and the corresponding $2 \times 2$ NNCT together with percentages
based on row and grand sums are provided in Table \ref{tab:NNCT-swamp}.
For example, for water tupelo as the base species and black gum as the NN species,
the cell count is 54 which is 26 \% of the 211 black gums (which is 54 \% of all 394 trees).
Observe that the percentages and Figure \ref{fig:SwampTrees} are suggestive of segregation for
all three tree species since the observed percentages of species with themselves as the NN are much larger
than the row percentages.

\begin{table}[ht]
\centering
\begin{tabular}{cc|cc|c}
\multicolumn{2}{c}{}& \multicolumn{2}{c}{NN species} & \\
\multicolumn{2}{c}{} & W.T. & B.G. &  sum  \\
\hline
& W.T. &    157 (74 \%) &   54 (26 \%) & 211 (54 \%)\\
\raisebox{1.25ex}[0pt]{base species}
& B.G. &    52  (28 \%) &  131 (72 \%) & 183 (46 \%) \\
\hline
& sum   &   209  (53 \%) &  185 (47 \%) &  394 (100 \%)\\
\end{tabular}
\caption{ \label{tab:NNCT-swamp}
The NNCT for swamp tree data and the corresponding percentages (in parentheses),
where the cell percentages are with respect to the row sums
and marginal percentages are with respect to the total size.
W.T. = water tupelos and B.G. = black gums.}
\end{table}

The locations of the tree species can be viewed a priori resulting
from different processes so the more appropriate null hypothesis is the CSR independence pattern.
Hence our inference will be a conditional one (see Section \ref{sec:QandR})
if I use the observed values of $Q$ and $R$.
I observe $Q=270$ and $R=236$ for this data set,
and the empirical estimates for these sample sizes
are $Q=249.68$ and $R=244.95$.
I present the tests statistics and the associated $p$-values for NNCT-tests
in Table \ref{tab:swamp-tree-NNCT-test-stat}.
Observe that the test statistics all decrease with the QR-adjustment,
however this decrease is not substantial to alter the conclusions.
Based on the NNCT-tests, I find that the segregation between both species is significant,
since all the tests considered yield significant $p$-values, and the
diagonal cells (i.e., cells $(1,1)$ and $(2,2)$) are larger than expected.

\begin{table}[ht]
\centering
\begin{tabular}{|c|c|c|c|}
\hline
\multicolumn{4}{|c|}{NNCT-test statistics and the associated $p$-values} \\
\multicolumn{4}{|c|}{for swamp tree data } \\
\hline
 $C$ & $\X_I^2$ & $\X_{II}^2$ & $\X_{III}^2$  \\
\hline
 52.72 & 52.08 & 52.14 & 52.66 \\
 ($<.0001$) & ($<.0001$) & ($<.0001$) & ($<.0001$) \\
\hline
\hline
 $C_{qr}$ & $\X_{I,qr}^2$ & $\X_{II,qr}^2$ & $\X_{III,qr}^2$  \\
\hline
 51.98 & 51.35 & 51.41 & 51.92\\
($<.0001$) & ($<.0001$) & ($<.0001$) & ($<.0001$) \\
\hline
\end{tabular}
\caption{ \label{tab:swamp-tree-NNCT-test-stat}
Test statistics and the associated $p$-values (in parentheses)
for NNCT-tests for the swamp tree data set.
$C$ stands for Dixon's overall test, $\X_I^2$, $\X_{II}^2$, and $\X_{III}^2$
stand for versions I, II, and III of the tests by \cite{ceyhan:corrected}.
$C_{qr}$, $\X_{I,qr}^2$, $\X_{II,qr}^2$, and $\X_{III,qr}^2$ are the
QR-adjusted versions of these tests.}
\end{table}

\subsection{Artificial Data Set}
\label{sec:arti-data}
\begin{figure}[ht]
\centering
\rotatebox{-90}{ \resizebox{3 in}{!}{\includegraphics{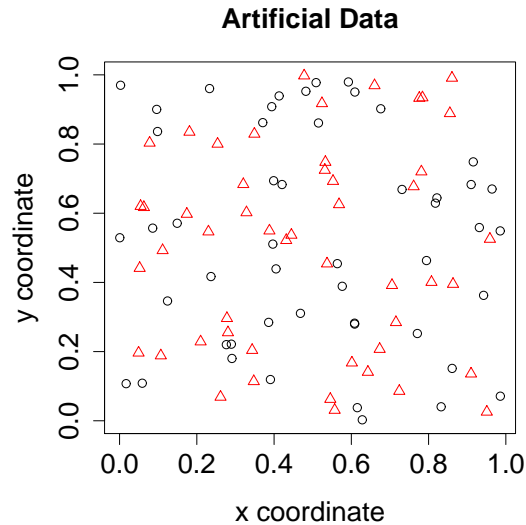} }}
\caption{
\label{fig:ArtiData}
The scatter plot of the locations of $X$ (circles $\circ$)
and $Y$ points (triangles $\triangle$) in the artificial data set.}
\end{figure}

In the swamp tree example,
although the test statistics for unadjusted and QR-adjusted versions
are different for Pielou's and Dixon's tests
and $p$-values for QR-adjusted versions are larger than unadjusted ones,
I have the same conclusion:
there is strong evidence for segregation of tree species.
Below, I present an artificial example,
a random sample of size 100 (with $50$ $X$-points and
$50$ $Y$-points uniformly generated on the unit square).
The question of interest is the spatial interaction between $X$ and $Y$ classes.
I plot the locations of the points
in Figure \ref{fig:ArtiData} and the corresponding NNCT together with
percentages are provided in Table \ref{tab:arti-data-NNCT-test-stat}.
Observe that the percentages are suggestive of mild segregation, with
equal degree for both classes.

\begin{table}[ht]
\centering
\begin{tabular}{cc|cc|c}
\multicolumn{2}{c}{}& \multicolumn{2}{c}{NN class} & \\
\multicolumn{2}{c}{} & $X$ & $Y$ &  sum  \\
\hline
& $X$ &    30 (60 \%) &   20 (40 \%) & 50 (50 \%)\\
\raisebox{1.25ex}[0pt]{base class}
& $Y$ &    19  (38 \%) &  31 (62 \%) & 50 (50 \%) \\
\hline
& sum   &   49  (49 \%) &  51 (51 \%) &  100 (100 \%)\\
\end{tabular}
\caption{ \label{tab:NNCT-arti}
The NNCT for the artificial data and the corresponding percentages (in parentheses),
where the cell percentages are with respect to the row sums
and marginal percentages are with respect to the total size.
}
\end{table}

The data is generated to resemble the CSR independence pattern,
so I assume the null pattern is CSR independence,
which implies that our inference will be a conditional one
if I use the observed values of $Q$ and $R$.
I observe $Q=70$ and $R=60$ for this data set,
and the empirical estimates for these sample sizes
are $Q=63.37$ and $R=62.17$.
I present the tests statistics and the associated $p$-values for NNCT-tests
in Table \ref{tab:arti-data-NNCT-test-stat}.
Observe that the test statistics all decrease with the QR-adjustment,
however this decrease is not substantial to alter the conclusions.
Based on the NNCT-tests, I find that the spatial interaction between $X$ and $Y$
is not significantly different from CSR independence.

In both examples although QR-adjustment did not change the conclusions,
it might make a difference if the pattern is a close call between CSR independence and
segregation/association.
That is, if a segregation test has a $p$-value about .05,
after the QR-adjustment, it might get to be significant or insignificant,
depending on the case.

\begin{table}[ht]
\centering
\begin{tabular}{|c|c|c|c|}
\hline
\multicolumn{4}{|c|}{NNCT-test statistics and the associated } \\
\multicolumn{4}{|c|}{$p$-values for the artificial data } \\
\hline
 $C$ & $\X_I^2$ & $\X_{II}^2$ & $\X_{III}^2$  \\
\hline
3.36 & 3.02 & 3.07 & 3.30 \\
(.1868) & (.0825) & (.2152) & (.0693) \\
\hline
\hline
 $C_{qr}$ & $\X_{I,qr}^2$ & $\X_{II,qr}^2$ & $\X_{III,qr}^2$  \\
\hline
3.32 & 2.97 & 3.04 & 3.25 \\
(.1906) & (.0846) & (.2192) & (.0713) \\
\hline
\end{tabular}
\caption{ \label{tab:arti-data-NNCT-test-stat}
Test statistics and the associated $p$-values (in parentheses)
for NNCT-tests for the artificial data set.
The notation for the tests is as in \ref{tab:swamp-tree-NNCT-test-stat}.}
\end{table}

\section{Discussion and Conclusions}
\label{sec:disc-conc}
In this article,
I discuss the effect of QR-adjustment on segregation or clustering tests
based on nearest neighbor contingency tables (NNCTs).
These tests include Dixon's overall test (\cite{dixon:1994}),
and the three new overall segregation tests introduced by (\cite{ceyhan:corrected}).
QR-adjustment is performed on these tests based on NNCTs (i.e., NNCT-tests)
when the null case is the CSR of two classes of points (i.e., CSR independence),
since under CSR independence, the NNCT-tests depend on
number of reflexive NNs (denoted by $R$) and the number of shared NNs (denoted by $Q$),
both of which depend on the allocation of the points.
When the observed values of $Q$ and $R$ are used,
the NNCT-tests are conditional tests, which might bias the results of the analysis.
Given the difficulty in calculating the expected values of $Q$ and $R$ under CSR independence,
I estimate them empirically based on extensive Monte Carlo simulations,
and substitute these estimates for expected values of $Q$ and $R$
(which is called the QR-adjustment in this article).

I compare the empirical sizes and power estimates of the NNCT-tests with extensive Monte Carlo simulations.
Based on the Monte Carlo analysis,
I find that QR-adjustment does not affect the empirical sizes of the tests.
Moreover, QR-adjustment does not have a substantial influence
on these NNCT-tests under the segregation or association alternatives.
Thus, one can use the QR-adjusted or the unadjusted versions of the NNCT-tests.



\end{document}